\documentclass[10pt,journal,compsoc]{IEEEtran}

\ifCLASSOPTIONcompsoc
  \usepackage[nocompress]{cite}
\else
  \usepackage{cite}
\fi

\IEEEoverridecommandlockouts
\usepackage{cite}
\usepackage{amsmath,amssymb,amsfonts}
\usepackage[hyphens]{url}
\usepackage{algorithm,algpseudocode}
\usepackage{mathtools}
\usepackage{graphicx}
\usepackage{textcomp}
\def\BibTeX{{\rm B\kern-.05em{\sc i\kern-.025em b}\kern-.08em
    T\kern-.1667em\lower.7ex\hbox{E}\kern-.125emX}}
\usepackage{multirow}
\usepackage{flushend}
\usepackage{stfloats}
\usepackage {booktabs}
\usepackage{cleveref}
\crefname{figure}{Fig.}{Figs.}
\crefname{table}{Table}{Tables}
\crefname{algorithm}{Algorithm}{Algorithms}
\crefname{section}{Section}{Sections}
\makeatletter
\let\MYcaption\@makecaption
\makeatother
\usepackage{subcaption}
\makeatletter
\let\@makecaption\MYcaption
\makeatother

\begin{document}

\title{ 
CSI-Inpainter: Enabling Visual Scene Recovery from CSI Time Sequences for Occlusion Removal}

\author{Cheng Chen, ~\IEEEmembership{Student Member,~IEEE},
        Shoki Ohta,~\IEEEmembership{Student Member,~IEEE},
        Takayuki Nishio,~\IEEEmembership{Senior Member,~IEEE},
        Mehdi Bennis,~\IEEEmembership{Fellow,~IEEE},
        Jihong Park,~\IEEEmembership{Senior Member,~IEEE}, and Mohamed Wahib
        
\IEEEcompsocitemizethanks{

\IEEEcompsocthanksitem Cheng Chen, Shoki Ohta, and Takayuki Nishio are with the School of Engineering, Tokyo Institute of Technology, Ookayama, Meguro-ku, Tokyo 152-8550, Japan.\protect\\
E-mail: \{chen.c.aj, ohta.s.ad\}@m.titech.ac.jp, nishio@ict.e.titech.ac.jp

\IEEEcompsocthanksitem Mehdi Bennis is with the Centre of Wireless Communications, University of Oulu, Oulu, 90014, Finland.\protect\\
E-mail: mehdi.bennis@oulu.fi

\IEEEcompsocthanksitem Jihong Park is with the School of Info Technology, Deakin University, Geelong Waurn Ponds, Australia.\protect\\
E-mail: jihong.park@deakin.edu.au

\IEEEcompsocthanksitem Mohamed Wahib and Cheng Chen are with RIKEN Center for Computational Science (RIKEN-CCS), Kobe, 650-0047, Japan.\protect\\
E-mail: mohamed.attia@riken.jp

}
\thanks{Manuscript received April 19, 2005; revised August 26, 2015.}}

\markboth{Journal of \LaTeX\ Class Files,~Vol.~14, No.~8, August~2015}%
{Shell \MakeLowercase{\textit{et al.}}: Bare Demo of IEEEtran.cls for Computer Society Journals}

\IEEEtitleabstractindextext{
\begin{abstract}

Introducing CSI-Inpainter, a pioneering approach for occlusion removal using Channel State Information (CSI) time sequences, this work propels the application of wireless signal processing into the realm of visual scene recovery. Departing from traditional occlusion removal, CSI-Inpainter leverages CSI data to construct and refine obscured visual elements in a scene, facilitating recovery independent of lighting conditions. Validated through comprehensive testing in both office and industrial environments, CSI-Inpainter demonstrates a robust capacity for discerning and reconstructing occluded segments, establishing a new frontier for obstacle removal. This first-of-its-kind framework offers a transformative perspective on how environmental visual information can be extracted from CSI, thereby broadening the scope for computer vision applications in everyday contexts.
\end{abstract}

}

\maketitle

\IEEEdisplaynontitleabstractindextext
\IEEEpeerreviewmaketitle

\IEEEraisesectionheading{
\section{Introduction}\label{sec:introduction}}

\IEEEPARstart{O}{bstacle} removal, encompassing both occlusion and image restoration in Computer Vision (CV), is pivotal for applications that demand a clear understanding of the environment, such as in surveillance systems, autonomous driving, and path prediction. The traditional scope of image inpainting—restoring missing or deteriorated portions of images—has been extended in recent years to include the removal of visual obstructions, leading to enhanced environmental perception \cite{maiti2021novel}, \cite{jam2021comprehensive}, \cite{Shu2016Inpainting}.

Deep Learning (DL) has significantly advanced the field of obstacle removal, surmounting many of the challenges inherent in traditional methods, \cite{angah2020removal}, \cite{Sharma2021Risk-Aware}, \cite{wang2020intelligent}, \cite{le2021physics}. Despite this progress, DL models often grapple with maintaining the fidelity of reconstructed images, especially under substantial occlusion  \cite{qin2021image}. Moreover, reliance on visual data alone can lead to inaccuracies when substantial portions of the image are obscured \cite{Liu_2018_ECCV}.

Radio Frequency (RF) signals offer a novel solution to these issues. Their ability to provide additional spatial information independent of visual data can significantly enhance the process of obstacle removal \cite{kato2021csi2image}. Building on our prior work with RF-Inpainter \cite{chen2022rf}, which leverages Received Signal Strength Indicator (RSSI) data for image inpainting, we introduce CSI-Inpainter. This novel approach utilizes CSI—a more granular and robust RF characteristic than RSSI—to augment the visual data, enabling a deeper understanding of the physical space for more accurate obstacle removal.

In this paper, we present CSI-Inpainter, a framework for occlusion and obstacle removal that leverages the power of CSI. The contributions of our work are manifold:

\begin{itemize}
\item We introduce the first method for visual scene recovery from CSI time sequences using Transformer architectures. This enables the generation of visual sequences independent of lighting conditions or even without conventional cameras.
\item We conduct comprehensive evaluations of CSI-Inpainter's performance across various real-world settings, including office and factory environments, demonstrating superior results over existing methods.
\item Our study delves into the impact of CSI data variations on image reconstruction, shedding light on future advancements in wireless sensing and CV. Through experimental analysis, we investigate the effects of leveraging CSI data from distinct sensors, amalgamating data from multiple sensors situated at varied locations, and modifying the temporal or frequency dimensions of CSI matrices. 
\end{itemize}

The remainder of this paper is organized as follows: Section 2 reviews the literature on DL-based image inpainting, DL-based obstacle removal, and multimodal image inpainting. Section 3 describes the CSI-Inpainter system, including the model architecture and training procedures. Section 4 presents experimental validations of CSI-Inpainter against other state-of-the-art methods. Section 5 examines the influence of CSI data variations on imaging performance. Finally, Section 6 concludes the paper and outlines directions for future research.

\section{Related Works}

\subsection{DL-based Image Inpainting}

Deep Learning has made significant progress in inpainting tasks. Compared to traditional algorithms, DL-based approaches have demonstrated improved effectiveness in capturing high-level semantics and producing superior results \cite{qin2021image}, \cite{JAM2021103147}, \cite{ZHANG202374}, \cite{xiang2023deep}.

CNNs and Generative Adversarial Networks (GANs) are the two most important neural networks in the study of DL-based inpainting methods, as indicated by several research contributions \cite{qin2021image}, \cite{ZHANG202374}, \cite{xiang2023deep}. GAN and CNN-based image inpainting methods have attracted much attention since the Context Encoder proposal by Pathak et al. \cite{pathak2016context}. Many researchers have proposed novel techniques to enhance the effectiveness of these methods. For example, some have introduced semantic attention layers to the GAN-based inpainting method, as proposed by Jiahui Yu \cite{yu2018generative}. Other researchers have proposed partial convolution to adapt the convolutional parameters to account for image breakage \cite{liu2018image}, \cite{xie2019image}.

In recent years, Transformer, a type of attention-based architecture, have demonstrated impressive results in Natural Language Processing (NLP) and high-level vision applications. Attention operators within Transformers excel in long-range modeling and dynamic weighting, which allows the model to borrow feature patches from distant uncorrupted patches to generate new patches for the corrupted regions, making them more suitable for image inpainting than CNNs and GANs. Several studies, including \cite{liu2019coherent}, \cite{wang2019musical}, and \cite{zeng2019learning}, have explored how contextual attention operators search the entire image to fill the missing regions. Ye Deng et al. proposed the contextual Transformer network, which models the affinity between uncorrupted and corrupted image regions and focuses on constructing affinity inside both uncorrupted and corrupted regions, resulting in better contextual information capture across multiple scales and more reliable inpainting \cite{deng2021learning}. Additionally, Wenbo Li et al. introduced the Mask-Aware Transformer, a large-hole inpainting method that combines the strengths of Transformers and convolutions by customizing an inpainting-oriented Transformer block where the attention module aggregates non-local information only from partial valid tokens, indicated by a dynamic mask \cite{li2022mat}.

\subsection{DL-based Obstacle Removal}

The realm of obstacle removal, a nuanced extension of DL-based image inpainting, is pivotal for applications that require the deletion of undesired elements and the reconstruction of occluded scenes, such as in content editing and surveillance scenarios. The advancements in deep learning have ushered in innovative architectures and methodologies aimed at the precise detection and elimination of obstructions.

Significant strides have been made in this field, exemplified by Zhan et al. \cite{Zhan_2020_CVPR}, who harnessed self-supervised learning for the removal of objects, thus capitalizing on the inherent statistics of images for enhanced restoration outcomes. Furthermore, Be{\v{s}}i{\'c} et al. focus on geometry-aware adversarial learning, leveraging the DynaFill framework to achieve spatial-temporal coherence in RGB-D video inpainting \cite{bevsic2022dynamic}. Another method, DDLP, extends deep latent particle models for unsupervised video prediction, employing Transformer-based dynamics for frame sequence reconstruction \cite{daniel2023ddlp}. 

Beyond these contributions, the field has seen the development of DL frameworks tailored for obstacle removal in diverse settings. Angah and Chen's adaptation of a U-Net architecture for image-to-image translation \cite{angah2020removal} showcases the model's efficacy in eradicating redundant objects and replenishing missing content within images from construction sites, highlighting its potential in context restoration. Sharma and Tokekar \cite{Sharma2021Risk-Aware} proposed a DL-based modular framework that applies image inpainting techniques to semantically segmented images. This approach facilitates navigation through occluded yet traversable areas, incorporating prediction uncertainties into risk-aware path planning.

Moreover, the challenge of identifying patch-based image inpainting forgeries has been addressed by Wang and Niu \cite{wang2020intelligent} through a Mask R-CNN approach. This method enhances the accuracy of distinguishing inpainted from non-inpainted regions, aiming to mitigate detection inaccuracies and missed areas. Additionally, Le and Samaras \cite{le2021physics} introduced a DL method for shadow removal that employs a linear illumination transformation model coupled with an inpainting network to refine results significantly, thereby reducing the mean absolute error for shadowed regions.

Regardless of the DL models used, current state-of-the-art obstacle removal techniques are limited in producing visually realistic and semantically reasonable scene, as they rely solely on residual information in the image to infer missing content, resulting in issues such as color discrepancies, blurriness in reconstructed images, and differences between the reconstructed and ground truth images \cite{qin2021image}, \cite{xiang2023deep}. Such shortcomings are unacceptable for practical applications like criminal investigations and traffic cameras.

\subsection{Multimodal Image Inpainting: A Novel Inspiration}

The concept of multimodal image inpainting extends the scope of inpainting by incorporating various data forms, such as text and RF signals, alongside visual information. Inspired by vision-language integration, models like DALL-E 2 \cite{ramesh2022hierarchical} and CLIP-Guided Diffusion \cite{ramesh2021zero} have demonstrated the capacity to generate or restore images based on textual descriptions, opening up new possibilities for creative and context-aware inpainting. However, the abstract nature of text and its limitation in conveying detailed visual information necessitate exploring other modalities. RF signals, particularly WiFi CSI and RSSI, emerge as promising alternatives, offering direct environmental attributes conducive to inpainting. Our investigation focuses on harnessing CSI for detailed environmental mapping and obstacle removal, distinguishing our work as a pioneering effort in multimodal inpainting that leverages the untapped potential of RF signals. By integrating CSI with DL models, we aim to overcome the limitations of traditional and text-based inpainting methods, offering enhanced solutions for complex and occluded scene recoveries.


\section{CSI-Inpainter: CSI-Guided Obstacle Removal via Multimodal Transformer}

\subsection{System Model}

\begin{figure}[t]
    \centering
    \includegraphics[scale=0.45]{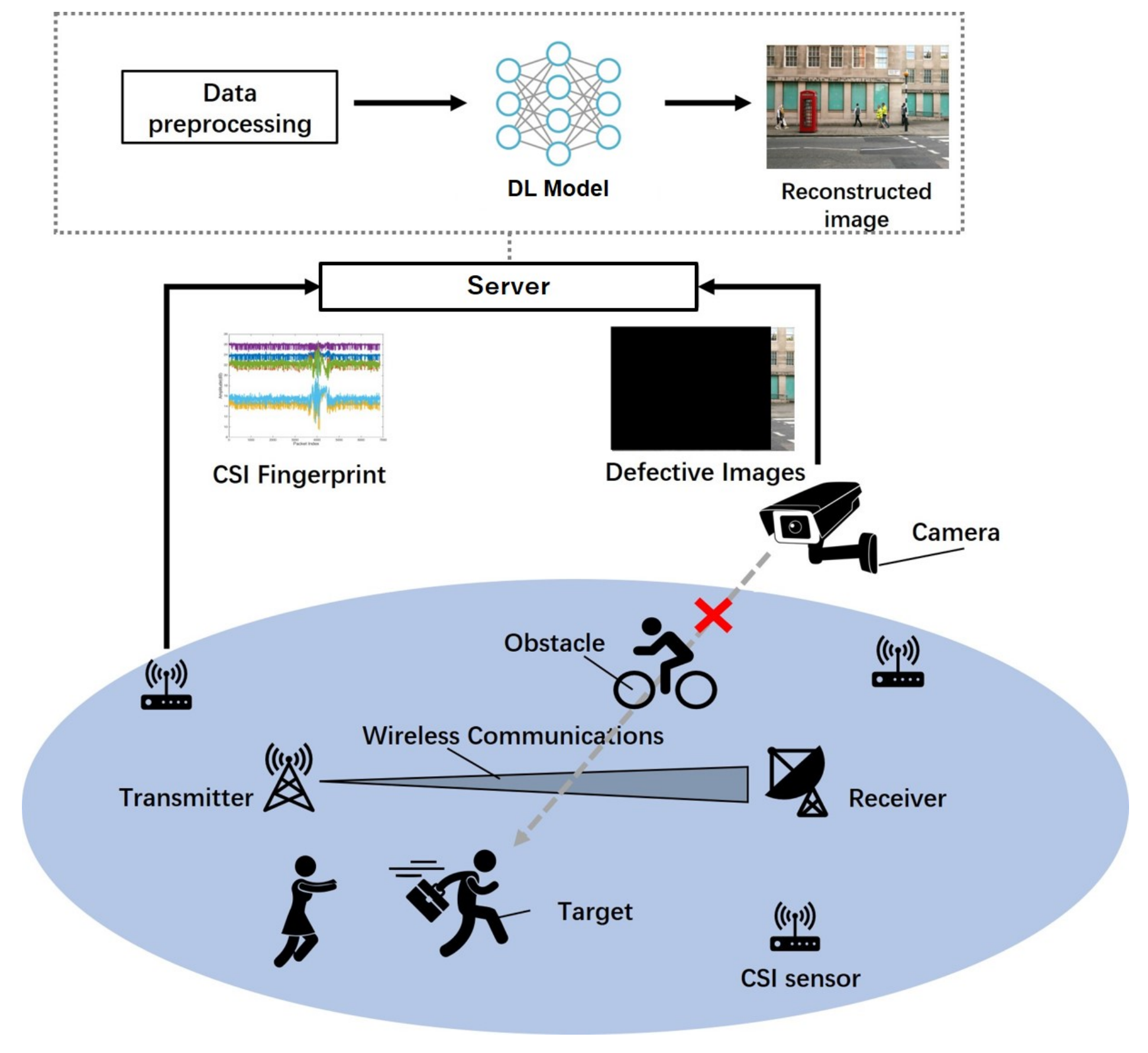}
    \caption{\label{fig:sys} The system model of CSI-Inpainter.}
\end{figure}

CSI-Inpainter combines camera and CSI sensor data to improve obstacle removal. As shown in \cref{fig:sys}, it consists of cameras, CSI sensors, a data processing module, and a deep neural network designed for removing obstacles from images.

Cameras provide RGB images, while CSI sensors collect wireless channel data. When images have missing sections due to obstructions, CSI data assists in predicting these areas, termed CSI-guided obstacle removal.

Data processing involves synchronizing and filtering image and CSI data, followed by feature extraction and PCA. The deep neural network uses a Transformer architecture to process defective video sequences and corresponding CSI sequences, restoring missing parts to match ground truth images.

\subsection{Data Collection and Preprocessing}
\subsubsection{Data Collection}

CSI is captured by WiFi transmitters using Long Training Symbols within the packet preamble. This complex-valued data represents the channel condition across different dimensions, such as space, frequency, and time. For image restoration tasks, we utilize only the amplitude information from the CSI, as it provides insights into environmental interactions with the WiFi signal.

The synchronization process is crucial to ensure each CSI data point is matched with an image frame. We synchronize clocks and standardize sampling rates for cameras and CSI sensors to achieve a consistent data collection rate of $10 fps$.

\subsubsection{Data Preprocessing}

\begin{figure}[t]
    \centering
    \includegraphics[scale=0.45]{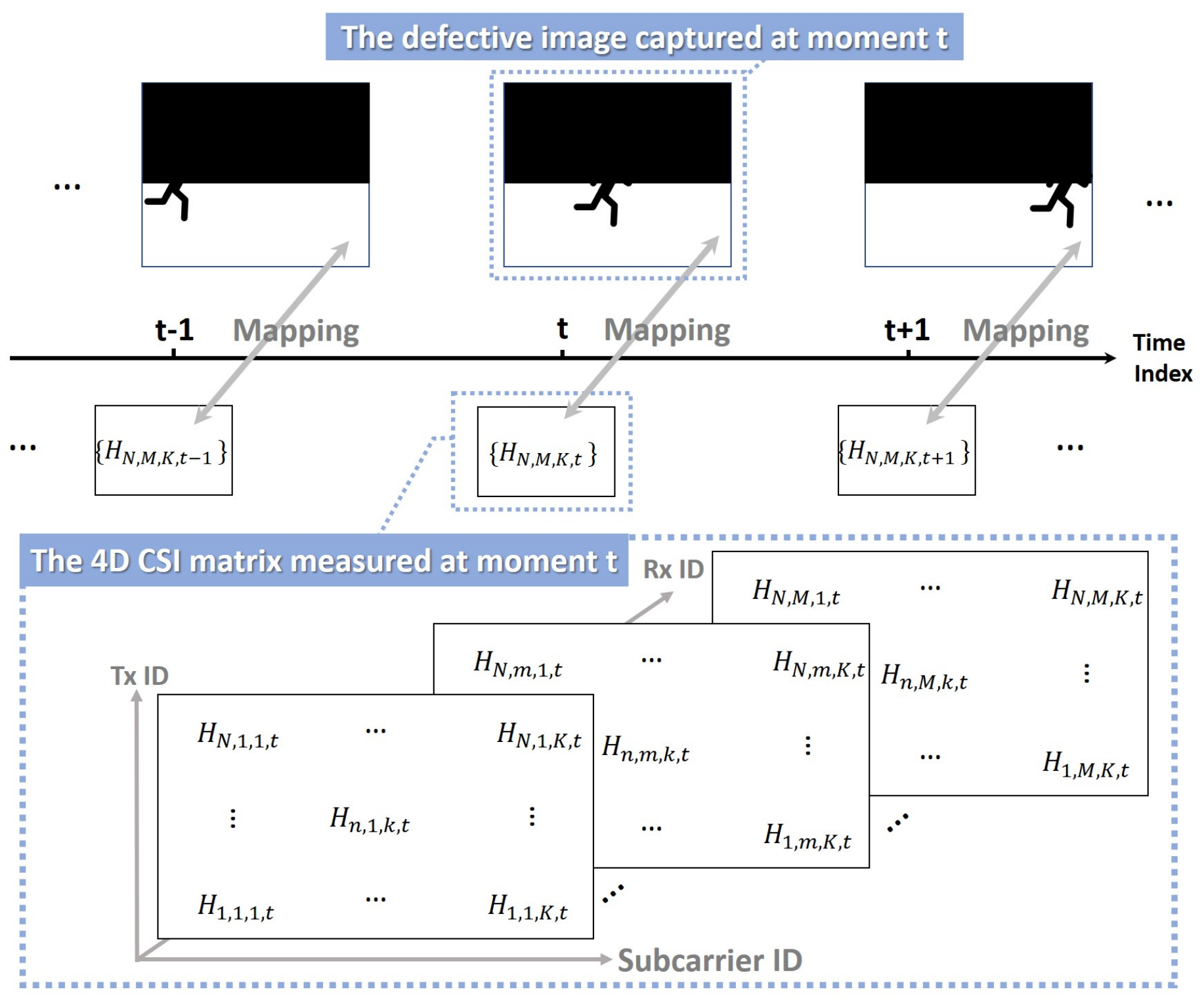}
    \caption{\label{fig:mapping} Mapping a 4D CSI matrix to the corresponding defective image.}
\end{figure}

Collected images are preprocessed to remove noise and resized to a standard dimension, while CSI data is cleaned to remove noise and irrelevant information. Isochronization ensures both data types are aligned temporally for integration into the deep learning framework. This involves matching data acquisition times to a set reference using a bisection search method, illustrated in \cref{fig:mapping}. Post-cleaning, we remove unnecessary subcarrier information from the CSI to improve data quality for the subsequent image restoration process.

\subsection{CSI-Inpainter Model Architecture}

\begin{figure*}[t]
    \centering
    \includegraphics[scale=0.35]{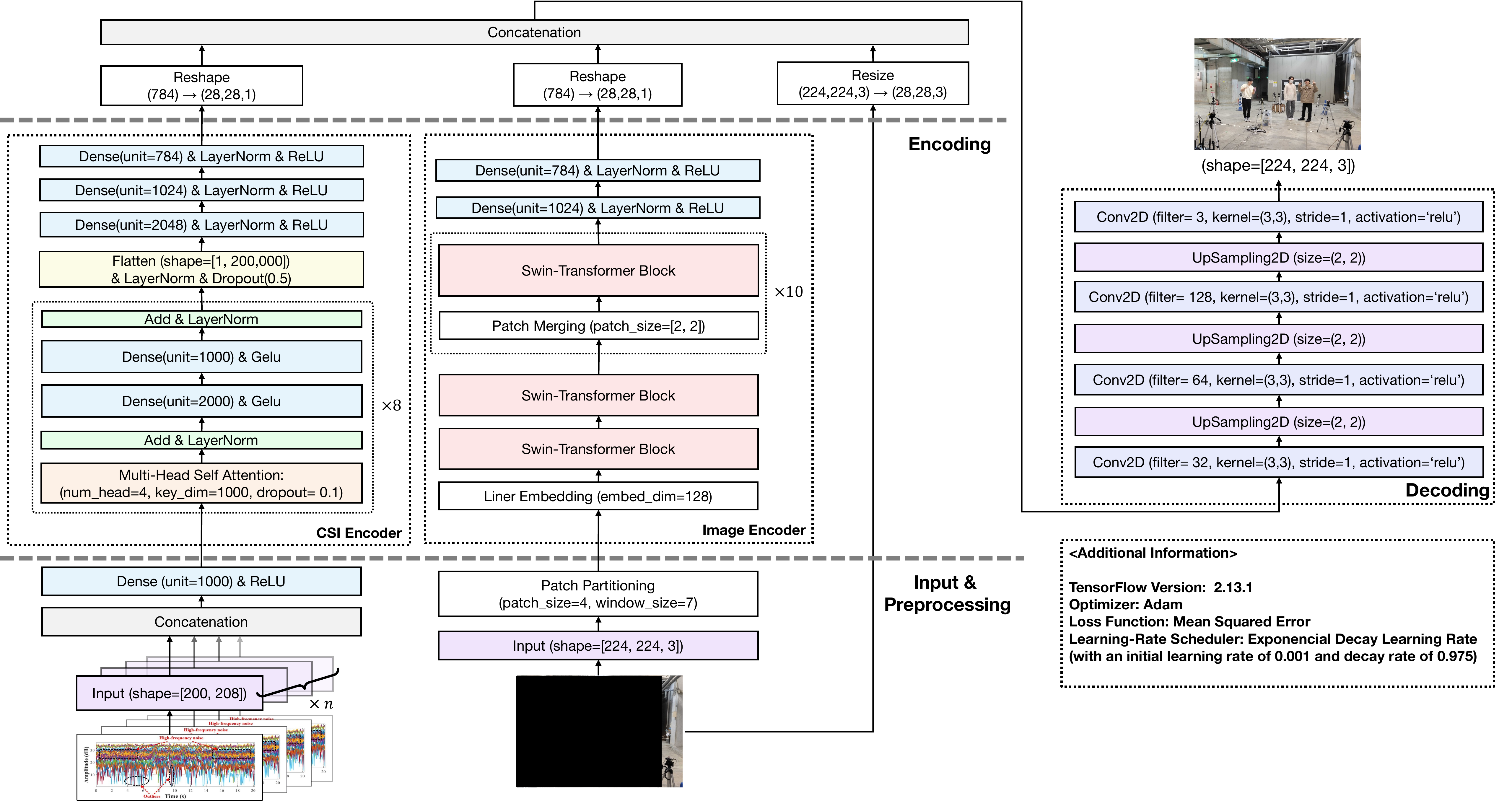}
    \caption{
    \label{fig:model}The model architecture of CSI-Inpainter.}
\end{figure*}

The CSI-Inpainter model integrates image and CSI data to remove obstacles from images. Its architecture consists of four parts: Input \& Preprocessing, Encoding, Aggregation, and Decoding, as shown in \cref{fig:model}.

\subsubsection{Input \& Preprocessing}

Image sequences undergo patch encoding to transform them into patch embeddings, which are then processed by the Transformer model. CSI data, representing signal attenuation, skips embedding and positional encoding due to its different data structure and is directly segmented into patches for further processing.

\subsubsection{Encoding}

The Encoding part features a CSI encoder based on conventional Transformer architecture, which extracts visual cues from CSI matrices. The image encoder is built on the Swin Transformer (Swin-TF), known for its efficiency and effectiveness in CV tasks \cite{liu2021swin}. It was pre-trained on ImageNet-22k and fine-tuned on ImageNet-1k to capture specific features for obstacle removal tasks.

\subsubsection{Aggregation}

This part combines the outputs of the image and CSI encoders, reducing their dimensionality and concatenating them with the input defective image sequence. This merged representation captures both visual and contextual information needed for obstacle removal.

\subsubsection{Decoding}

The final part generates the restored image sequence, using convolutional and upsampling layers. Skip connections between input and output layers ensure the preservation of image content and enhance the realism of the output.

Through this architecture, CSI-Inpainter effectively learns to restore images by understanding the intricate relationships between visual and CSI data, leading to improved obstacle removal results.

\subsection{Training and Prediction Procedures}

Training involves learning from paired image and CSI matrix datasets. The model predicts missing parts using CSI and image data, applying learned patterns to generate obstacle removal results.

\section{Performance Evaluation}

This section provides a detailed description of the experiments used to evaluate the performance of CSI-Inpainter in two quintessential environments: office and factory settings. 

\subsection{Setup}
\subsubsection{Settings for office experiment}

\begin{table}[t]
\caption{Experimental equipment}
\label{tab:exp_equip}
\centering
\begin{tabular}{cc} \toprule
Receiver            & NETGEAR Nighthawk X10 \\
Transmitter         & NETGEAR Nighthawk X10 \\
Wireless LAN standard & IEEE 802.11ac                \\
Channel             & 36                    \\
Bandwidth           & 80\,MHz               \\ \midrule
CSI sensor          & Raspberry Pi 4 model B       \\
CSI sensor firmware & Nexmon CSI~\cite{nexmon_csi}            \\
CSI measurement rate & 500\,Hz \\ \midrule
Camera 1,2          & RealSense L515        \\
Camera 3            & RealSense D435       \\ \bottomrule
\end{tabular}
\end{table}

\begin{figure}[t]
    \begin{tabular}{cc}

     \begin{minipage}{\linewidth}
       \centering
       \includegraphics[scale=0.42]{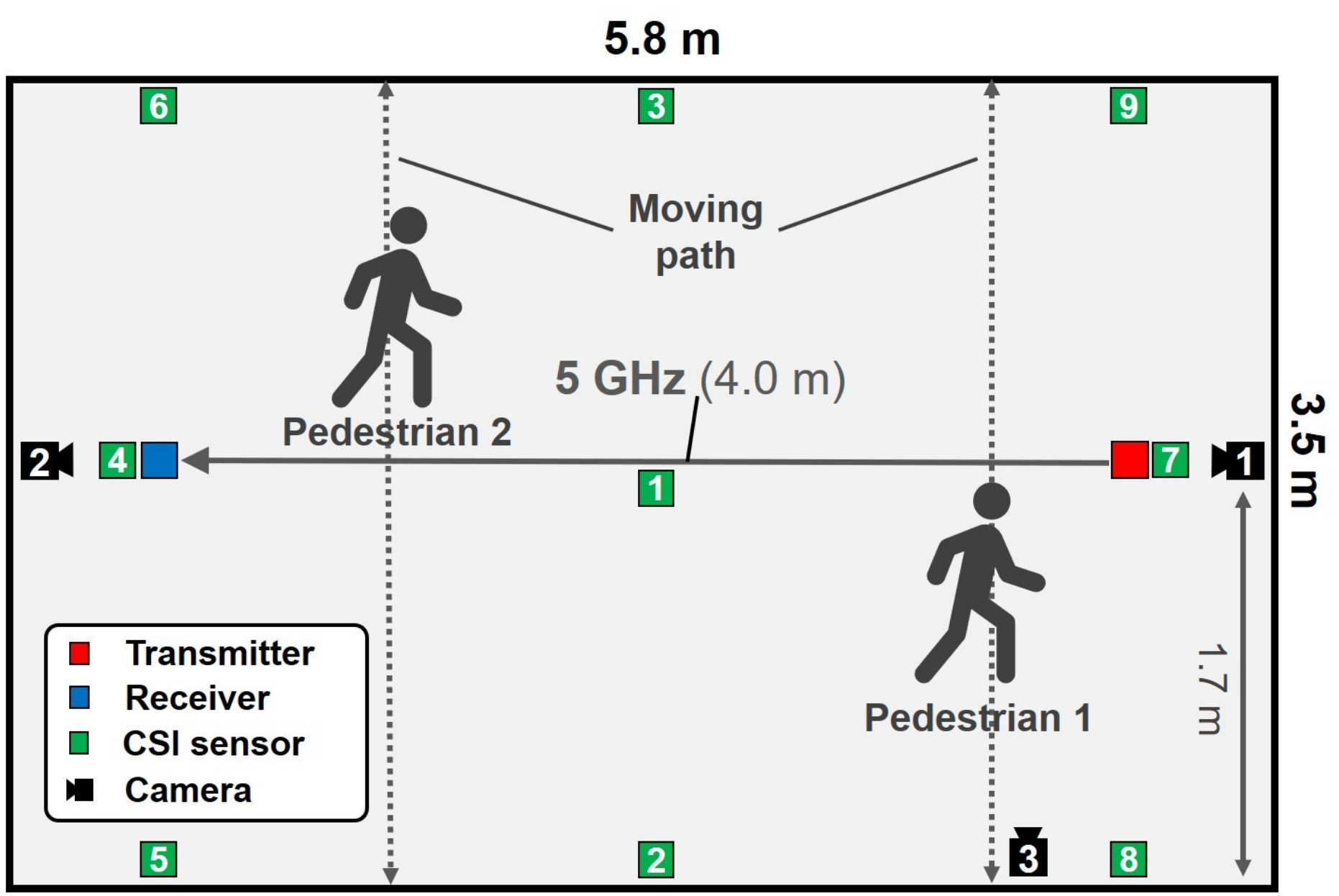}
       \subcaption{\label{fig:setup} The setup (each number represents the ID of each CSI sensor).}
     \end{minipage}\\

      \begin{minipage}{\linewidth}
        \centering
        \includegraphics[scale=0.4]{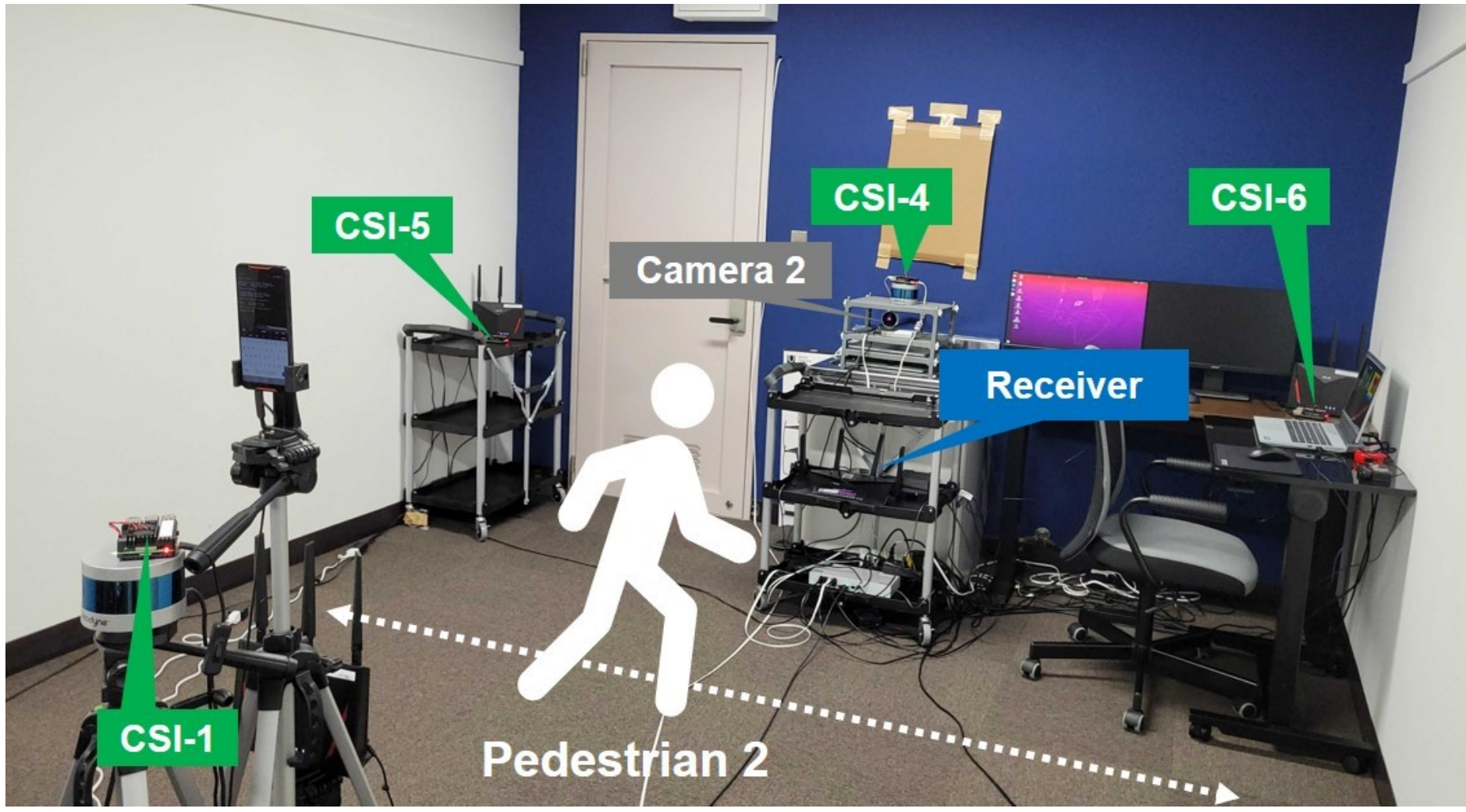}
        \subcaption{\label{fig:envir01} Snapshot 1.}
      \end{minipage}\\

      \begin{minipage}{\linewidth}
        \centering
        \includegraphics[scale=0.4]{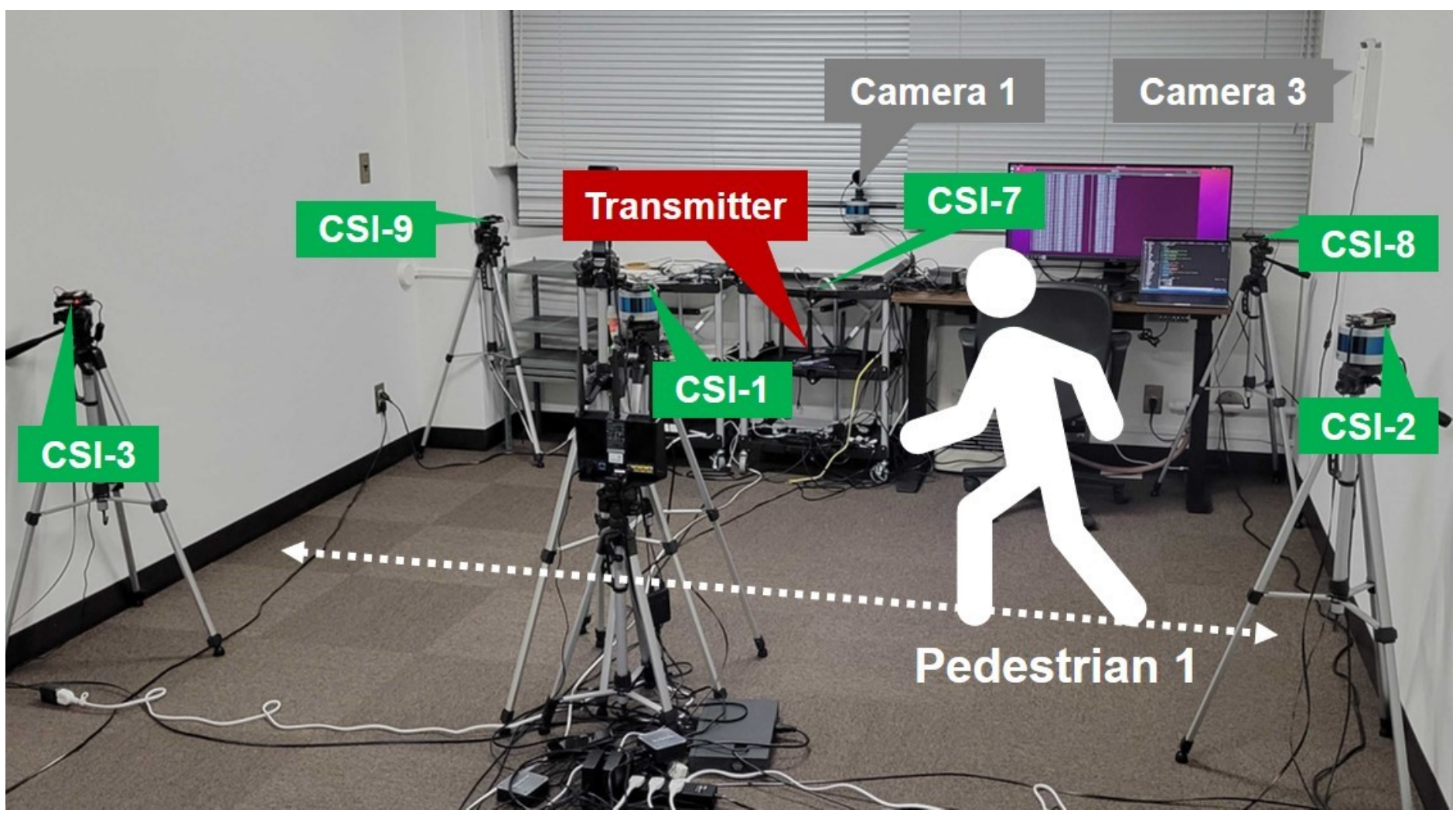}
        \subcaption{\label{fig:envir02} Snapshot 2.}
      \end{minipage}
      
    \end{tabular}
    \caption{\label{fig:offi_envir} Configuration for the office experiment.}
\end{figure}

The first experiment was conducted in an office, where the line-of-sight path of a 5\,GHz band IEEE 802.11ac wireless LAN connection was periodically obstructed by two pedestrians. The equipment used in the experiment and the routes taken by the pedestrians are depicted in \cref{fig:setup}. Snapshots of the environment are shown in \cref{fig:envir01} and \cref{fig:envir02}. To generate traffic, wireless LAN devices were installed at both ends of the room, and iperf was used as the traffic generator \cite{tirumala1999iperf}. To capture the wireless LAN signal and obtain CSI, nine CSI sensors were strategically placed in the experimental environment. These sensors were also responsible for collecting RSSI data, facilitating the comparison between CSI and RSSI for image inpainting. The movement of pedestrians caused variations in the radio propagation environment, leading to changes in both CSI and RSSI values. The experimental environment, including the pedestrians' movements, was captured by three RGB cameras. The equipment used in the experiment is listed in \cref{tab:exp_equip}. The CSI data was acquired using custom firmware, Nexmon CSI, on a Raspberry Pi. It is important to note that the Raspberry Pi with Nexmon CSI was not used for wireless communication but instead obtained CSI information by sniffing the signals from the receiver in monitor mode. For ease of reference, each CSI sensor is labeled as CSI-$n$ according to its ID number.

The experiment was conducted over 30 minutes. Data collection and preprocessing were performed following the method described in Section 3.2. This resulted in temporally continuous sequences of RGB images captured from three different angles, nine corresponding sequences of CSI matrices from different sensors, and nine sequences of RSSI values. Each image sequence consisted of 18,000 RGB images with dimensions $768 \times 1280$, each CSI sequence contained 18,000 matrices, and each RSSI sequence comprised 18,000 consecutive RSSI values.

\subsubsection{Settings for factory experiment}

\begin{figure*}[h]
    \centering
    \begin{minipage}{0.4\linewidth}
        \centering
        \includegraphics[scale=0.4]{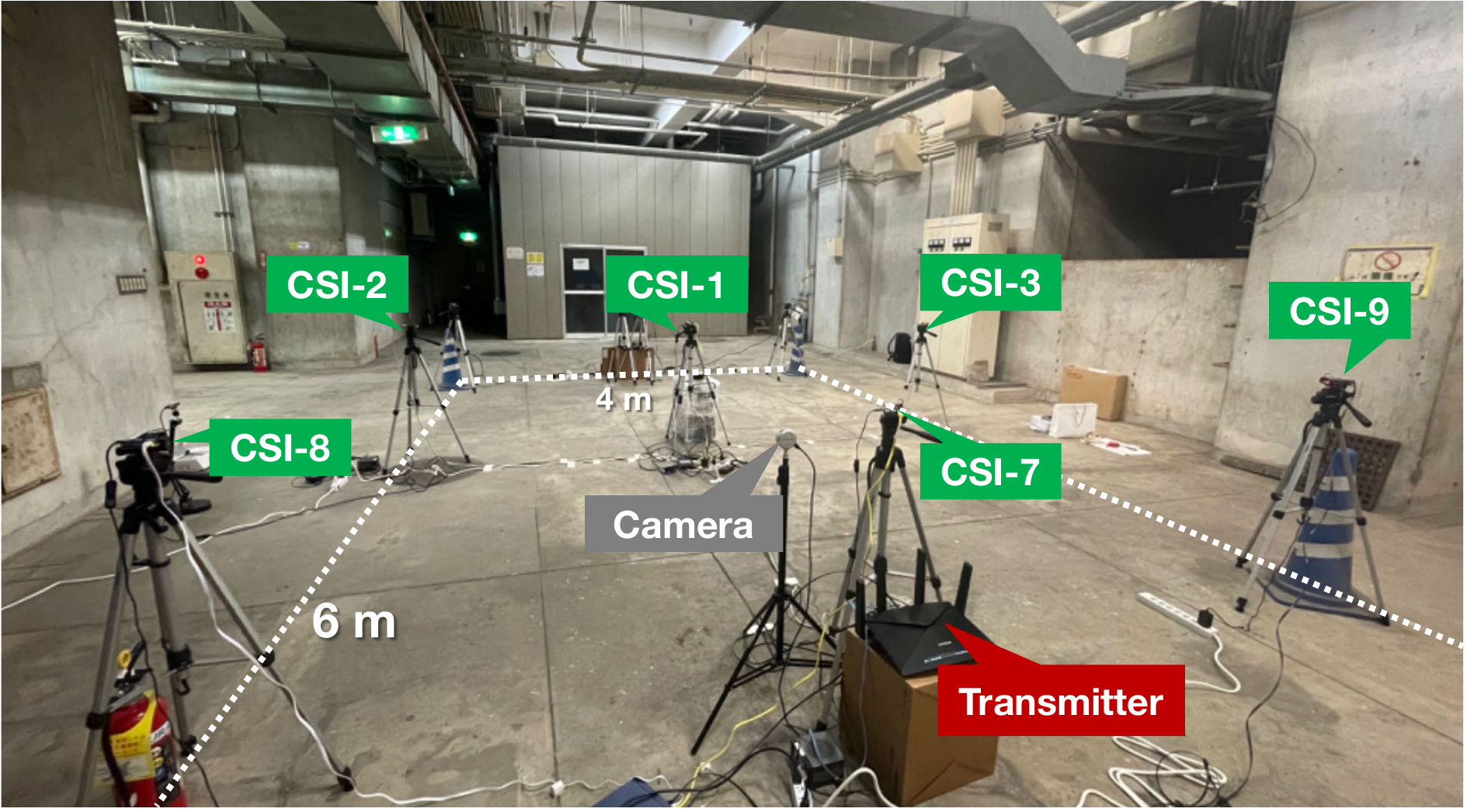}
        \subcaption{\label{fig:envir01} Snapshot 1}
    \end{minipage}
    \hfill 
    \begin{minipage}{0.4\linewidth}
        \centering
        \includegraphics[scale=0.4]{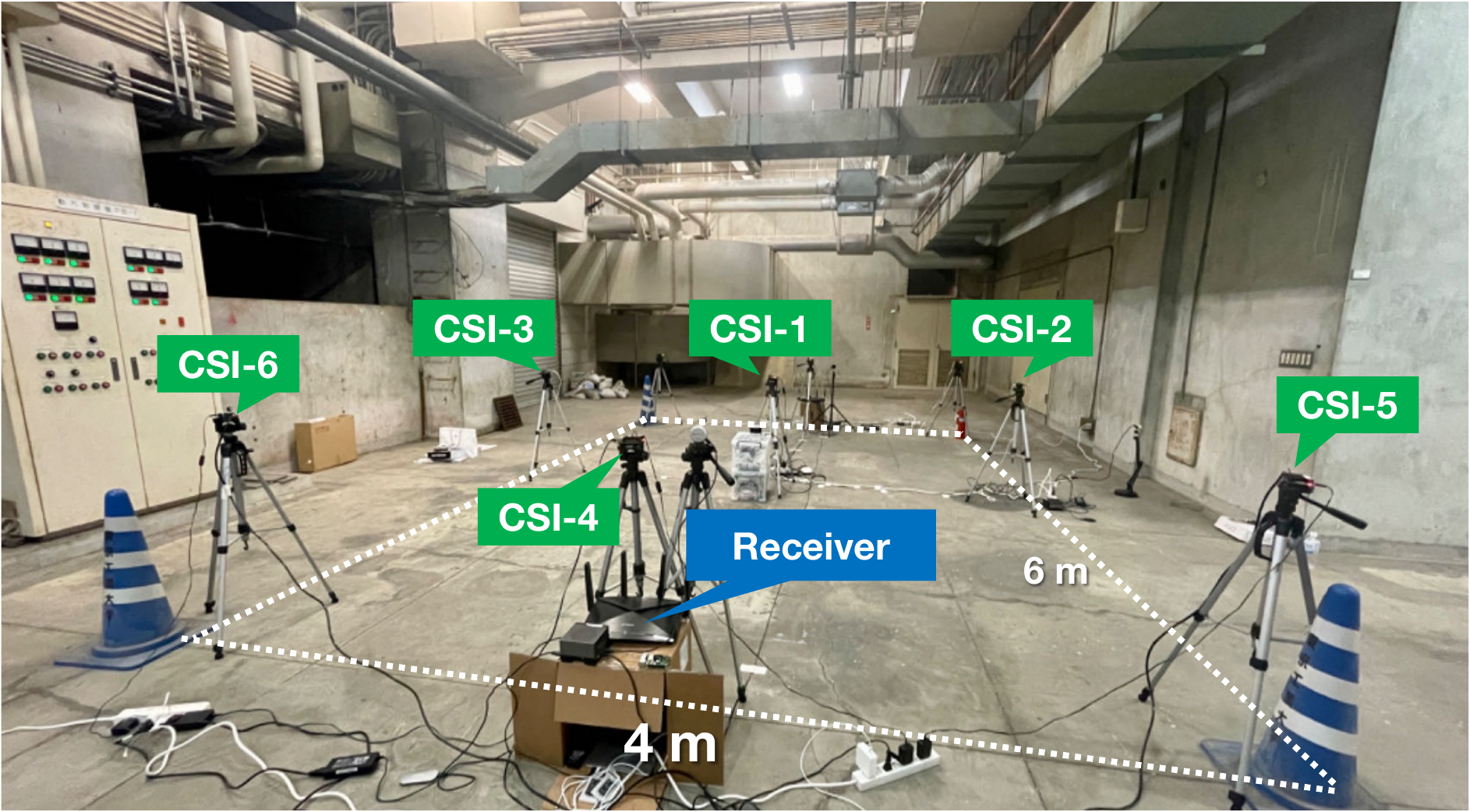}
        \subcaption{\label{fig:envir02} Snapshot 2}
    \end{minipage}\\ 
    \begin{minipage}{0.4\linewidth}
        \centering
        \includegraphics[scale=0.36]{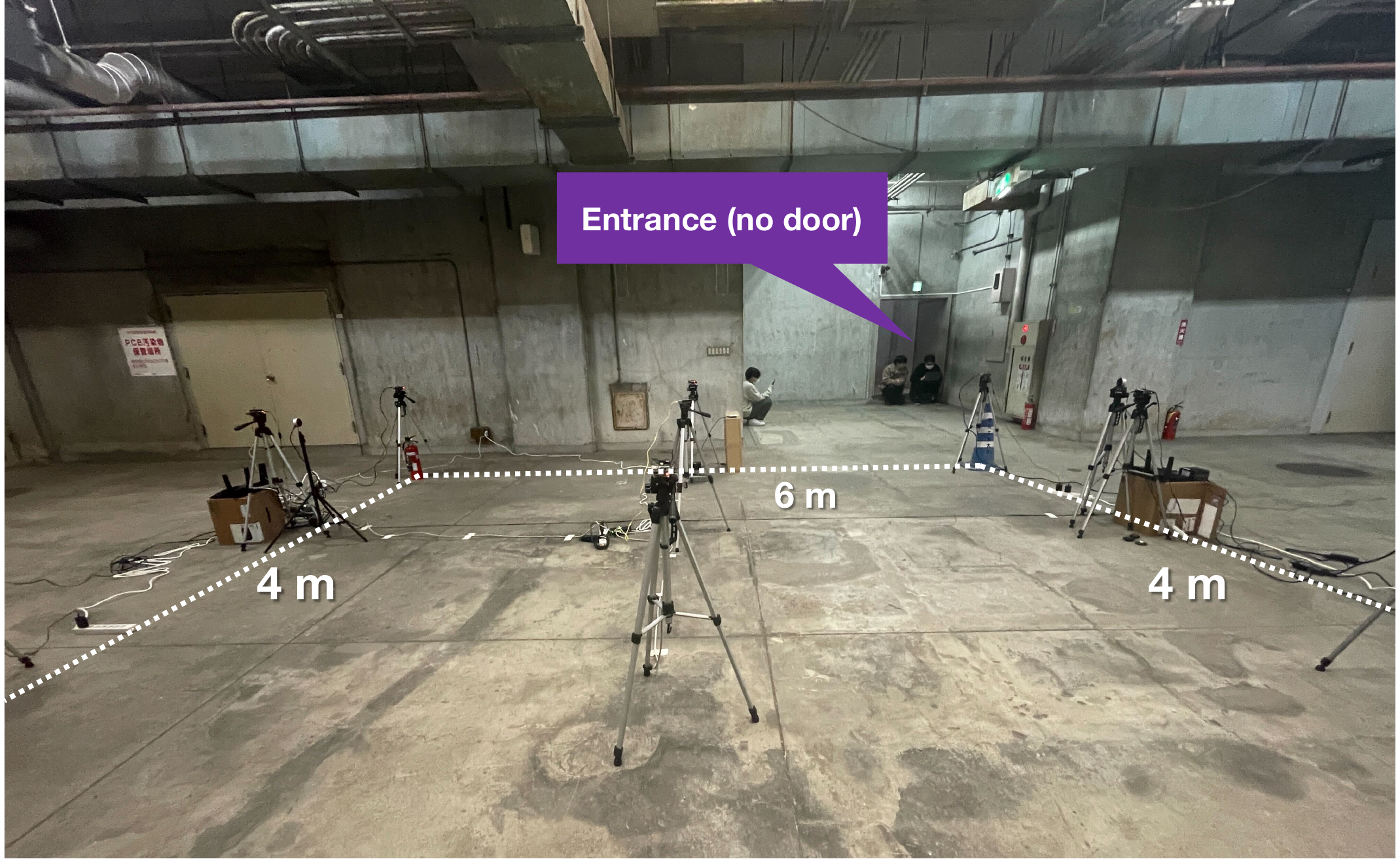}
        \subcaption{\label{fig:envir03} Snapshot 3}
    \end{minipage}
    \hfill 
    \begin{minipage}{0.4\linewidth}
        \centering
        \includegraphics[scale=0.36]{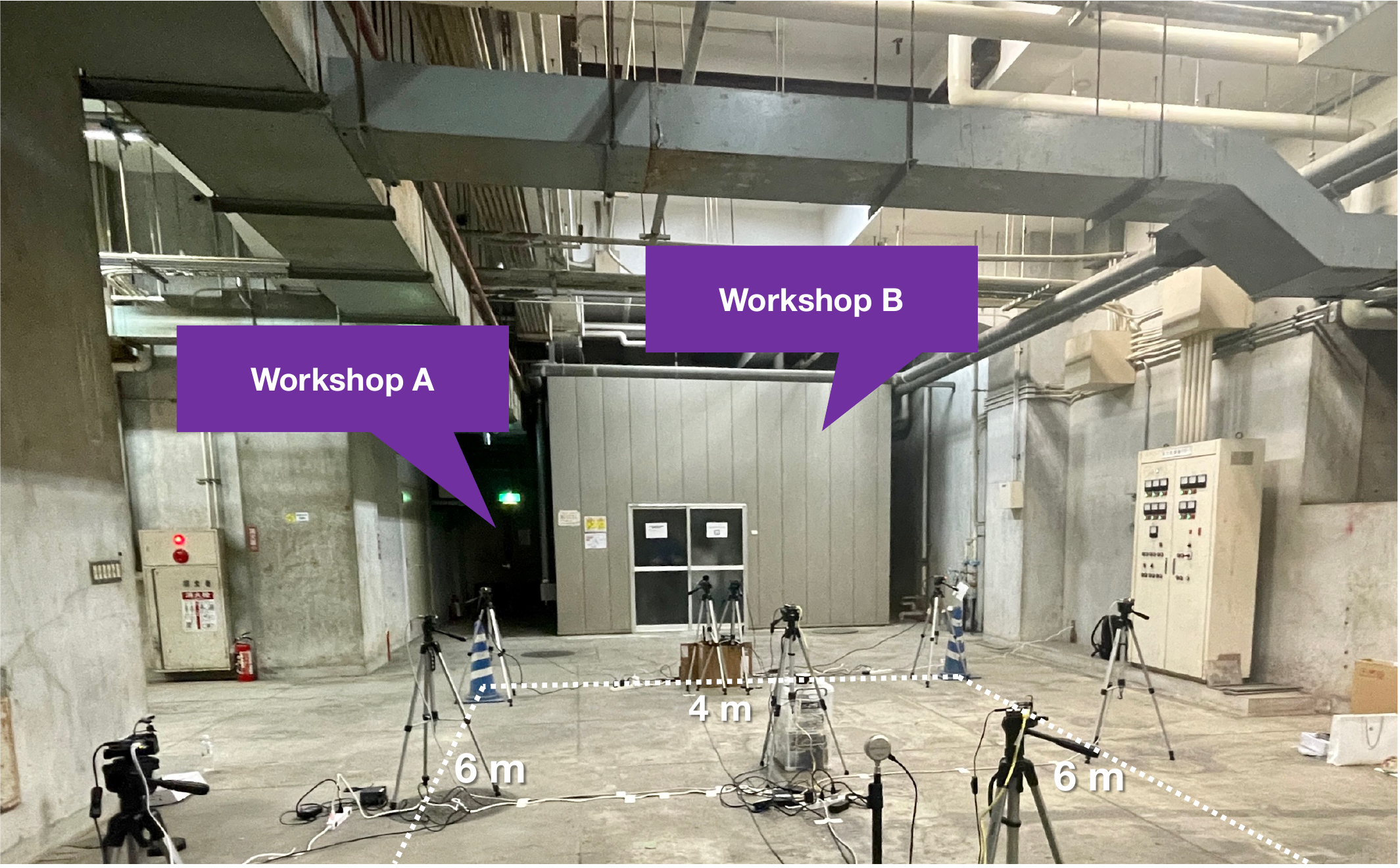}
        \subcaption{\label{fig:envir04} Snapshot 4}
    \end{minipage}
    \caption{\label{fig:fac_envir} Configuration for the factory workshop.}
\end{figure*}

To further validate the feasibility and robustness of CSI-Inpainter, we conducted complementary experiments in a more challenging and complex factory workshop (shown in \cref{fig:fac_envir} ). In this environment, the accuracy of CSI may suffer from interference of noise caused by the machine operation from other adjacent workshops, or sudden invasion caused by pedestrians unrelated to our experiments, etc. Besides, the room is much more spacious and not closed (the workshop entrance has no door), such that the amount of reflected Wi-Fi signals captured by sensors may largely degrade due to the limited transmission distance and semi-open-air environment.

To guarantee good results, we try to control the change in the experimental setting as little as possible. Therefore, we move all the devices from the office environment into the rectangle factory space with identical size ($6m \times 4m$) marked out by the white point lines shown in \cref{fig:fac_envir}. 

We continuously collected CSI-Images pairs datasets for the following three single-person scenarios. Please note that the colors of clothes and faces of these pedestrians are different. Also, to make sure the results is reliable, we ask the pedestrian in two adjacent cases to move along a counter-direction: 

\begin{itemize}
  \item \textbf{Scenario 1 (S1):} Pedestrian 1 (P1) walk around the experimental area (marked out with white point-lines) clockwise for 10 min.
  \item \textbf{Scenario 2 (S2):} Pedestrian 2 (P2) walked around the experimental area counterclockwise for another 10 min.
  \item \textbf{Scenario 3 (S3):} Pedestrian 3 (P3) walked around the experimental area clockwise for the last 10 min.
\end{itemize}

Building upon the single-person scenarios (S1 to S3), we further investigated multi-person scenarios involving three (S4 and S5) and five individuals (S6), capturing the data in a continuous sequence. The experimental scenarios are detailed as follows:

\begin{itemize}
  \item \textbf{Scenario 4 (S4):} A group consisting of Pedestrian 1, 2, and 3 (P1, P2, and P3) navigated the delineated area in a clockwise direction for 10 minutes.
  \item \textbf{Scenario 5 (S5):} A subsequent group comprising Pedestrian 1, 4, and 5 (P1, P4, and P5) traversed the same space counter-clockwise for another 10 minutes.
  \item \textbf{Scenario 6 (S6):} All five participants (P1 through P5) independently followed straight-line paths within the designated area for 20 minutes.
\end{itemize}

\subsection{Baselines \& Metrics}

RF-Inpainter, a state-of-the-art RF-based multimodal inpainting approach, serves as the primary baseline for comparison in this study. Alongside, we examine the performance of image-only CSI-Inpainter which utilizes Swin-TF as its backbone. To further assess the effectiveness of our approach, we incorporated comparisons against two notable conventional inpainting methods: the Hourglass Attention Network (HAN) by Deng Ye et al.\cite{deng2022hourglass} and Partial Convolutions (PConv) by Guilin Liu et al. \cite{Liu_2018_ECCV}. HAN, leveraging an attention-based architecture, demonstrates improved handling of images with invalid pixels due to its novel Laplace attention mechanism. PConv method mitigates common inpainting artifacts by conditioning convolutions on only valid pixels, showing significant benefits for irregular masks.

To ensure an objective evaluation of the obstacle removal performance of the respective methods, we utilize two widely adopted metrics for assessing image quality: the mean Peak Signal-to-Noise Ratio (PSNR) and the mean Structural Similarity Index (SSIM). These metrics facilitate a quantitative measurement of the inpainting quality by calculating the PSNR and SSIM values for all the inpainted test images.

Initially, we conduct a comprehensive comparative analysis of the multimodal obstacle-removal capabilities demonstrated by both CSI-Inpainter and RF-Inpainter. Next, we investigate specific scenarios where only data from vision is available. Finally, we evaluate images obtained solely from our respective wifi signals, which highlight the substantial amount of visual information embedded in the CSI matrix, as opposed to the RSSI value sequence.

\subsection{Comprehensive Performance Evaluation of CSI-Inpainter in Office Environments}

\begin{figure*}[h]
    \centering
    \includegraphics[scale=0.95]{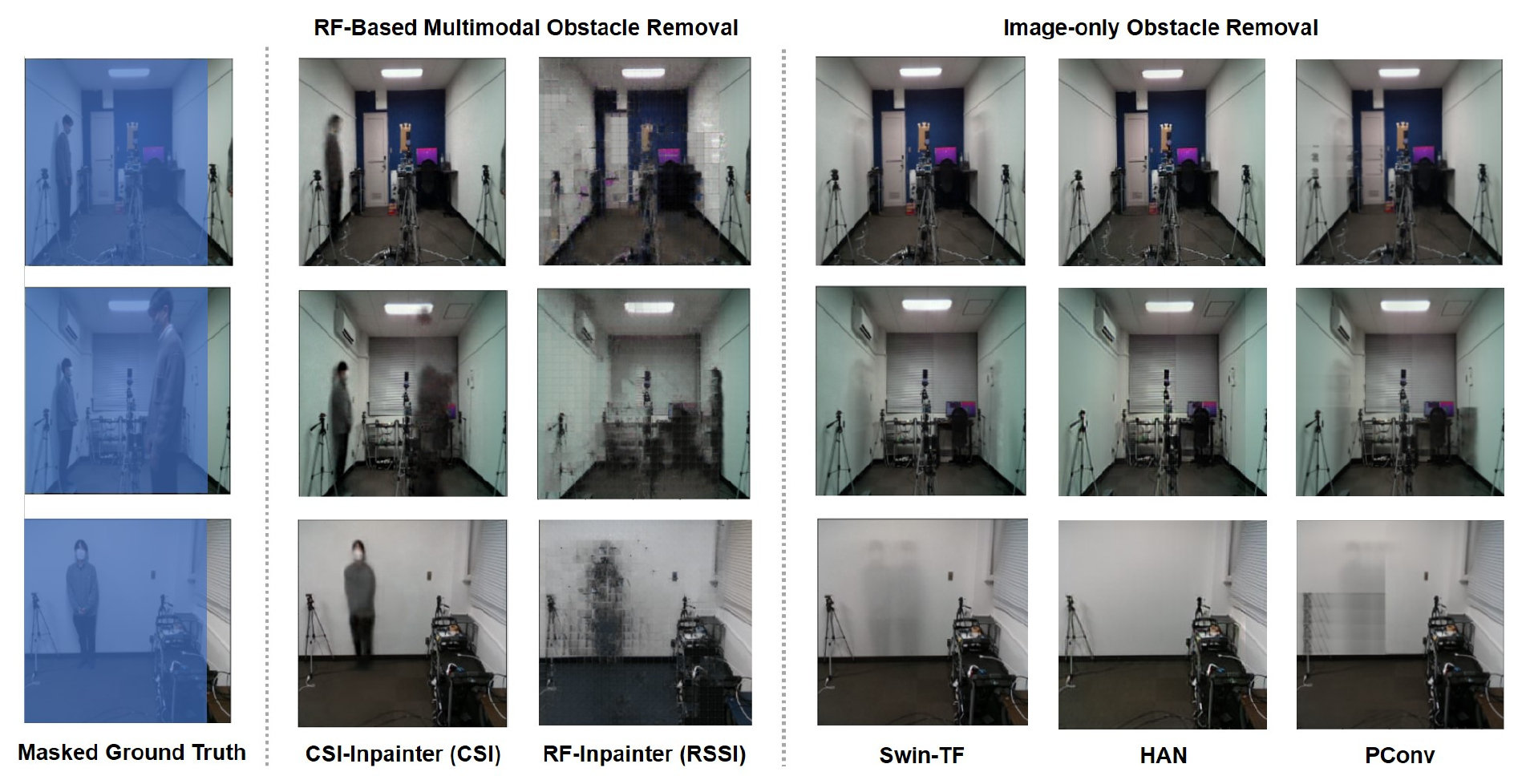}
    \caption{
    \label{fig:obstacle_removal_office} Sample obstacle removal results in office environments for Camera 1, Camera 2, and Camera 3, shown in the first through third rows, respectively. Multimodal CSI-Inpainter consistently excels in removing obstructions and restoring scenes across all scenarios.}
\end{figure*}

\begin{figure*}[h]
    \centering
    \includegraphics[scale=0.95]{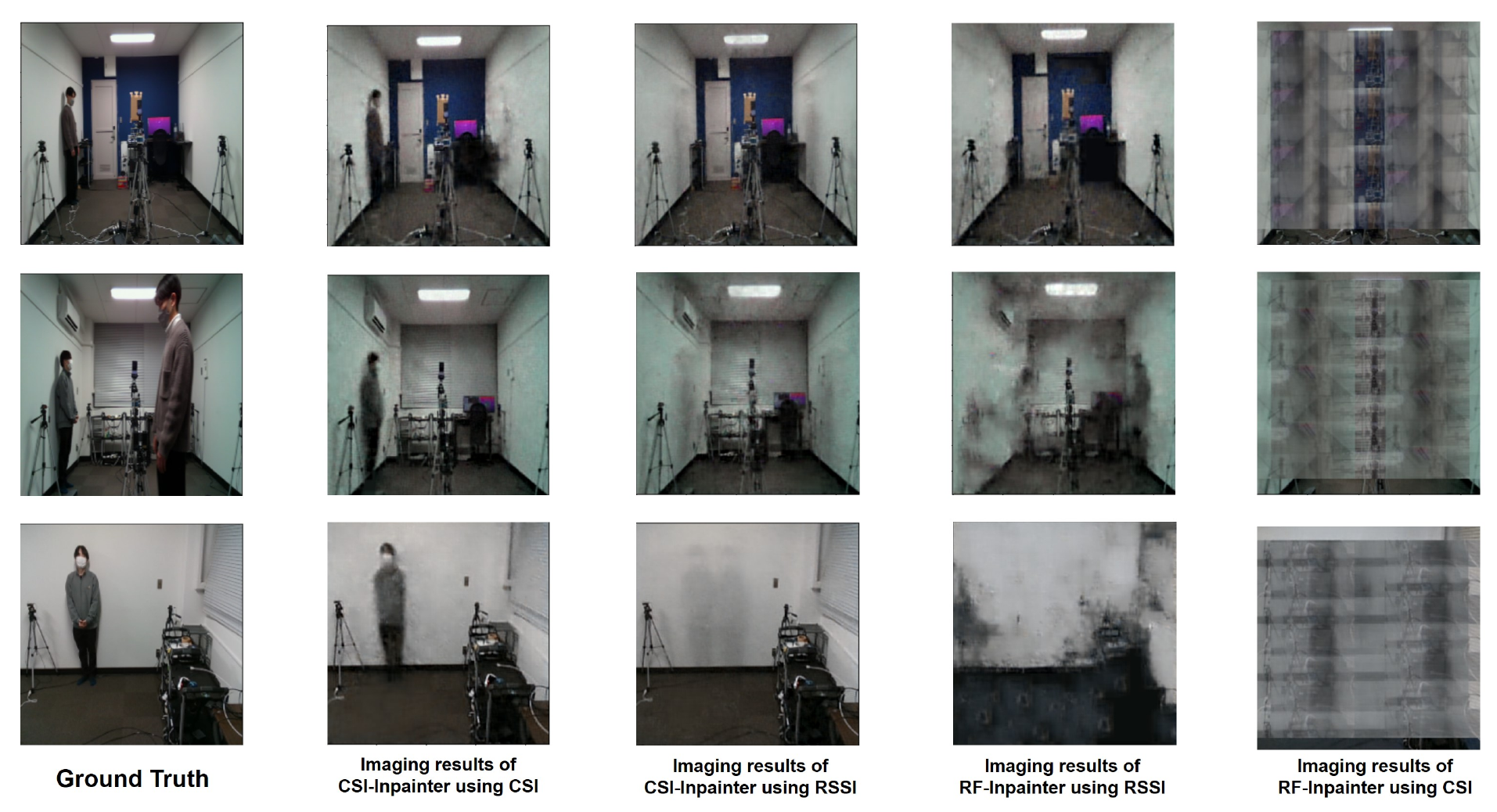}
    \caption{\label{fig:imaging} Sample imaging results for Camera 1, Camera 2, and Camera 3 are shown in the first through third rows, respectively. It can be seen that feeding CSI matrix sequence into CSI-Inpainter is the optimal solution for RF-based imaging as it yields the most complete images.}
\end{figure*}

\begin{figure*}[h]
    \centering
    \includegraphics[scale=0.9]{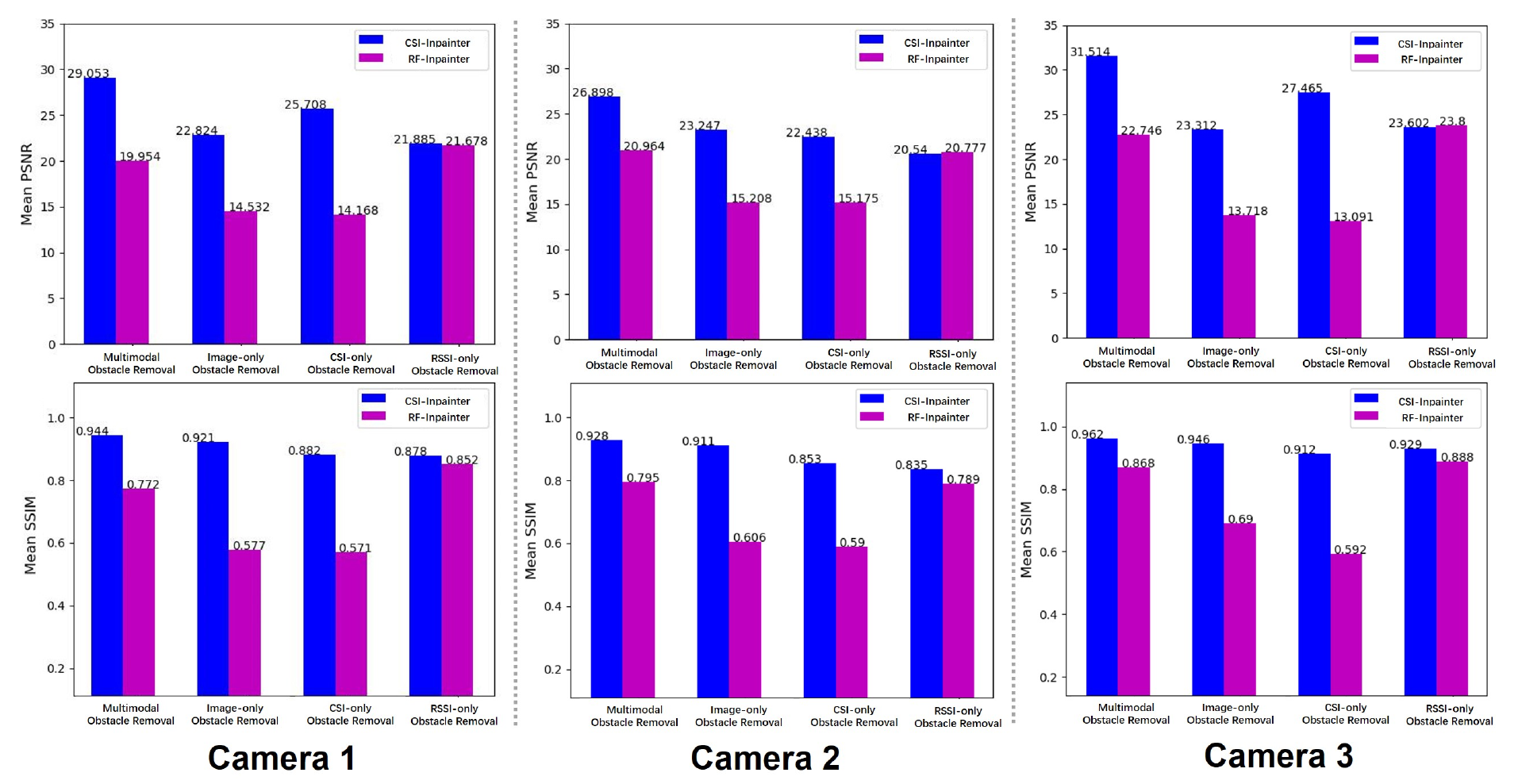}
    \caption{
    \label{fig:obj-result} Objective metrics for evaluating the performance of each method on each camera dataset. They showcase that multimodal CSI-Inpainter achieves maximum PSNR and SSIM scores in all scenarios.}
\end{figure*}

\subsubsection{Multimodal Obstacle Removal}

Multimodal obstacle removal necessitates the use of both defective images and CSI data. In the experiments we focused on instances with extensive occlusions, successfully demonstrating our method's efficacy even when over 90\% of the image content is obscured. This capability suggests broader applicability, as previously validated in \cite{chen2022rf}.

To ensure alignment with the local office environment, CSI and RSSI selection from sensors corresponded with the camera's perspective. For instance, using Camera 3's dataset, CSI data was drawn from sensors CSI-1, CSI-3, CSI-7, and CSI-9, reinforcing the integrative strength of multimodal data in superior obstacle removal outcomes.

For optimal results with RF-Inpainter, we relied on RSSI data, while CSI data was pivotal in enhancing the obstacle removal performance of CSI-Inpainter. Our findings, as depicted in \cref{fig:obstacle_removal_office} and \cref{fig:obj-result}, confirm that CSI-Inpainter leverages CSI data more effectively than RF-Inpainter, which fares better with RSSI data. Notably, employing CSI from multiple sensors afforded a qualitative edge, aligning with our prior research indicating that richer RF data quantity can improve obstacle removal quality.

\subsubsection{Obstacle Removal Using Limited Visual Information}

Even with minimal residual visual information, CSI-Inpainter demonstrates a notable performance advantage over the other baselines, adeptly restoring all impaired areas, as shown in \cref{fig:obstacle_removal_office}. While RF-Inpainter also shows commendable performance by utilizing RSSI data, it underscores the challenge faced by traditional inpainting techniques (Swin-TF, HAN, PConv) that rely purely on residual visual information for dynamic figure reconstruction. This comparative analysis not only showcases the efficacy of CSI-Inpainter but also enriches the discourse on leveraging CSI for enhanced imaging outcomes in complex environments.

\subsubsection{RF-based Obstacle Removal}

Our exploration of various model and RF data combinations aimed to determine the most effective method for RF-based obstacle removal. \cref{fig:imaging} demonstrates that CSI-Inpainter, employing CSI matrix sequences, excels at producing clear and complete images. This efficacy is evidenced by the highest mean PSNR and SSIM scores, as illustrated in \cref{fig:obj-result}. Notably, these results are achieved without relying on traditional visual inputs. In contrast, RF-Inpainter's performance is significantly reduced when using CSI, as shown in \cref{fig:imaging}. This shortfall is due to RF-Inpainter's architecture being optimized for RSSI data. RSSI, being one-dimensional, contains less information than the complex four-dimensional amplitude information characteristic of CSI. Consequently, RF-Inpainter is not adequately equipped to handle CSI data's detailed and multidimensional nature. Despite a decline in performance with RSSI data, CSI-Inpainter still surpasses RF-Inpainter in visual information extraction, as depicted in both \cref{fig:imaging} and \cref{fig:obj-result}. The reduced effectiveness with RSSI can be attributed to CSI-Inpainter's design, which is specifically built for the nuanced, multidimensional CSI data. Operating with less comprehensive RSSI data means the model's full potential is not utilized, leading to diminished effectiveness.


\subsection{Extended Validation of CSI-Inpainter in Industrial Environments}

\begin{figure*}[h]
    \centering
    \begin{minipage}{\linewidth}
        \centering
        \includegraphics[scale=0.57]{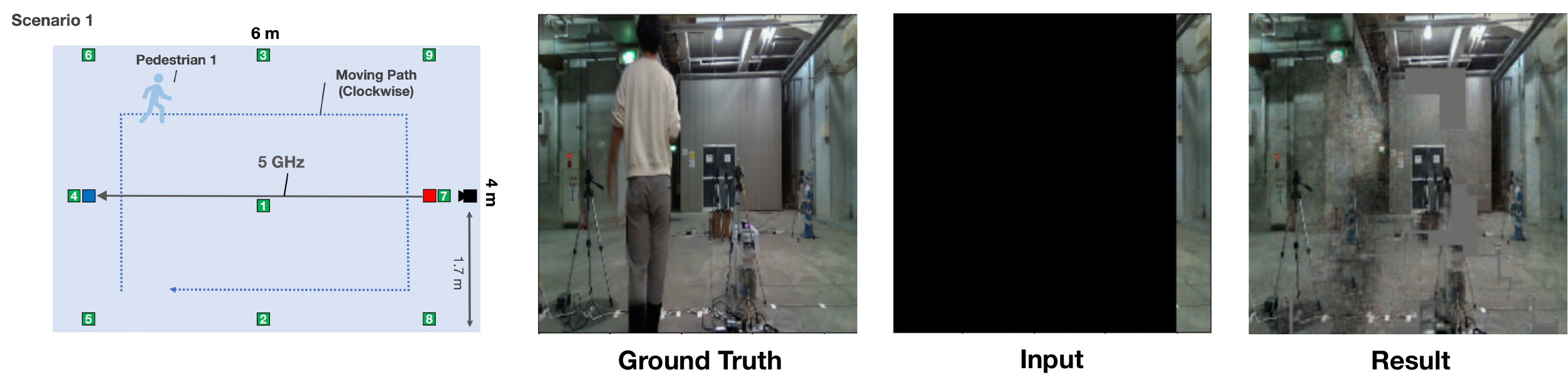}
        \subcaption{Result of Scenario 1.}
        \label{fig:factory_result_01}
    \end{minipage}
    \\ 
    
    \begin{minipage}{\linewidth}
        \centering
        \includegraphics[scale=0.57]{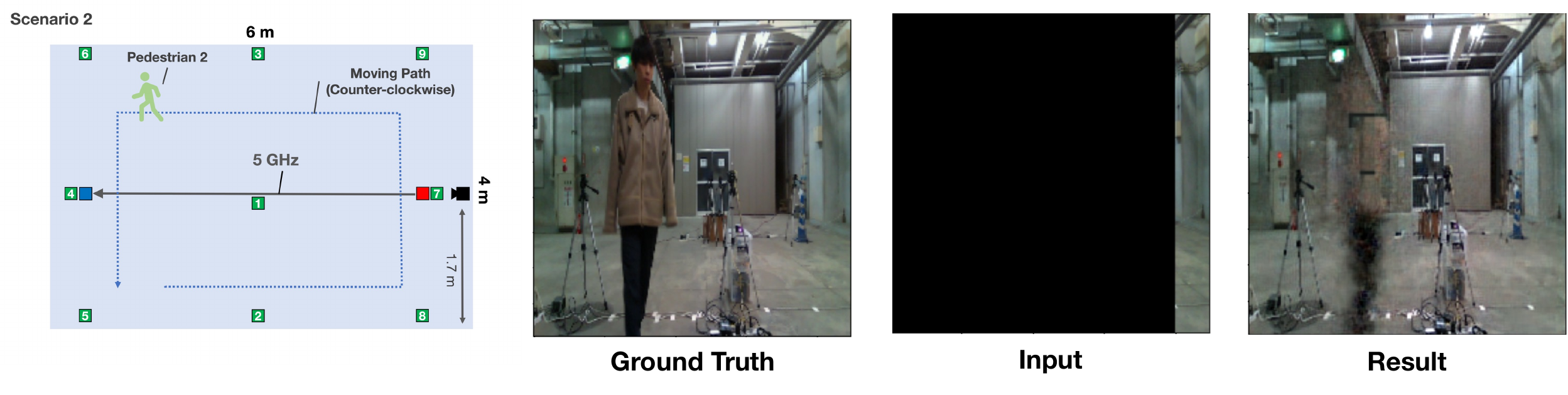}
        \subcaption{Result of Scenario 2.}
        \label{fig:factory_result_02}
    \end{minipage}
    \\ 

    \begin{minipage}{\linewidth}
        \centering
        \includegraphics[scale=0.57]{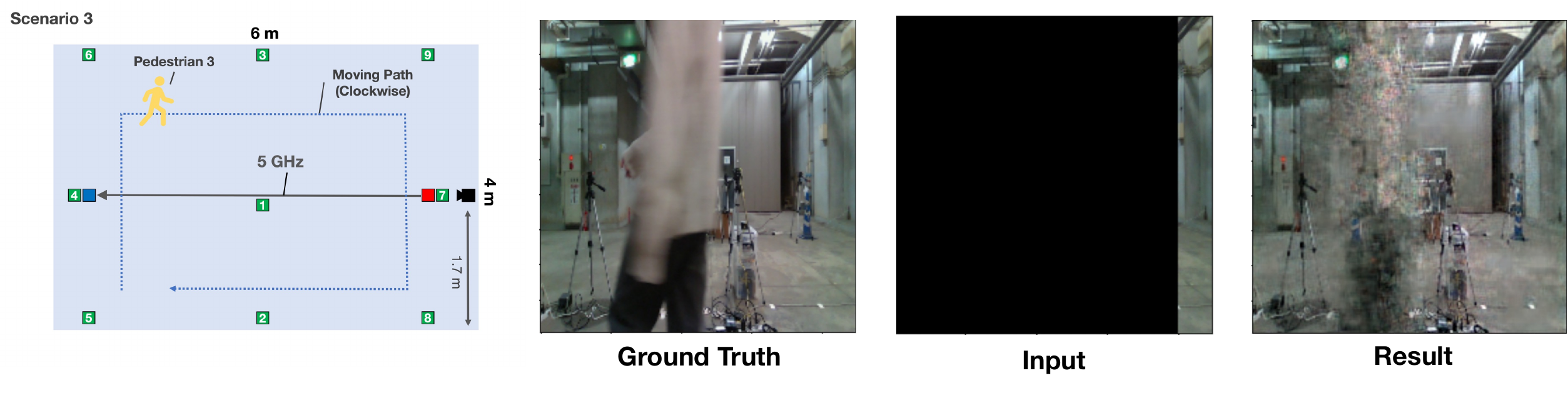}
        \subcaption{Result of Scenario 3.}
        \label{fig:factory_result_03}
    \end{minipage}
    \caption{
    \label{fig:Single_Pedestrian_results} Obstacle Removal for Single-Pedestrian Scenarios.}
\end{figure*}

\begin{figure}[h]
    \centering

    \begin{minipage}{\linewidth}
        \centering
        \includegraphics[scale=0.45]{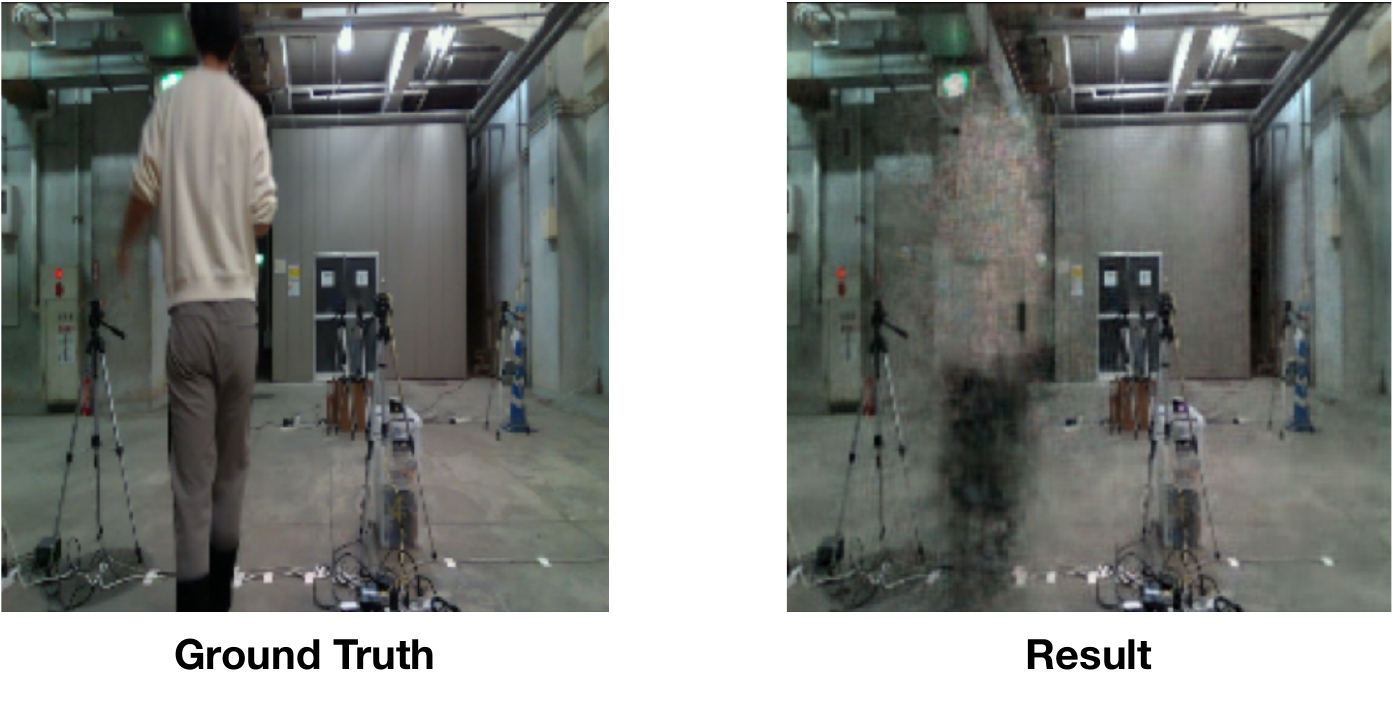}
        \subcaption{S1 Revisited: Accurate Recovery of P1}
        \label{fig:scenario1_revisited}
    \end{minipage}
    \hfill
    
    \begin{minipage}{\linewidth}
        \centering
        \includegraphics[scale=0.45]{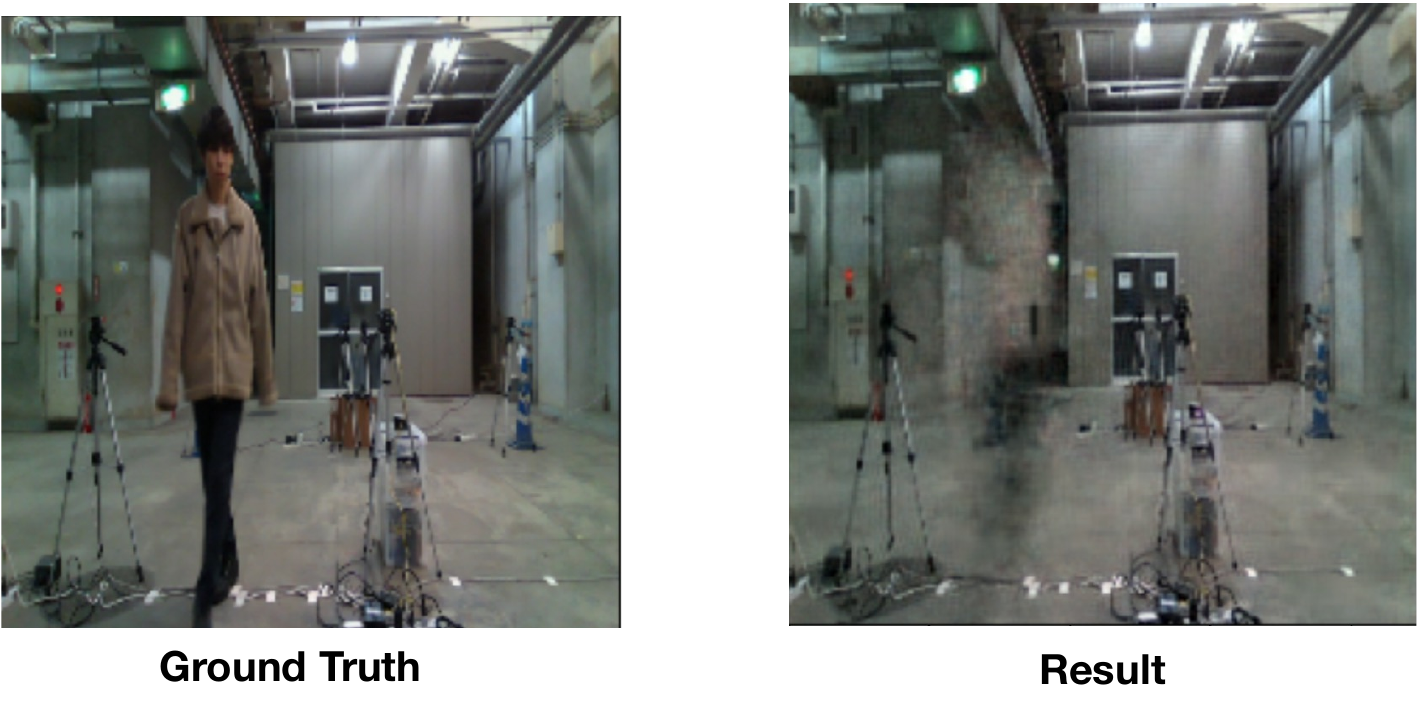}
        \subcaption{S2 Revisited: Location Accuracy with Color Discrepancy}
        \label{fig:scenario2_revisited}
    \end{minipage}

    \caption{ 
    \label{fig:revisited_scenarios} Re-evaluation of S1 and S2 after Fine-Tuning on S3 Dataset.}
\end{figure}

\begin{figure*}[h]
    \centering
    
    \begin{minipage}{\linewidth}
        \centering
        \includegraphics[scale=0.55]{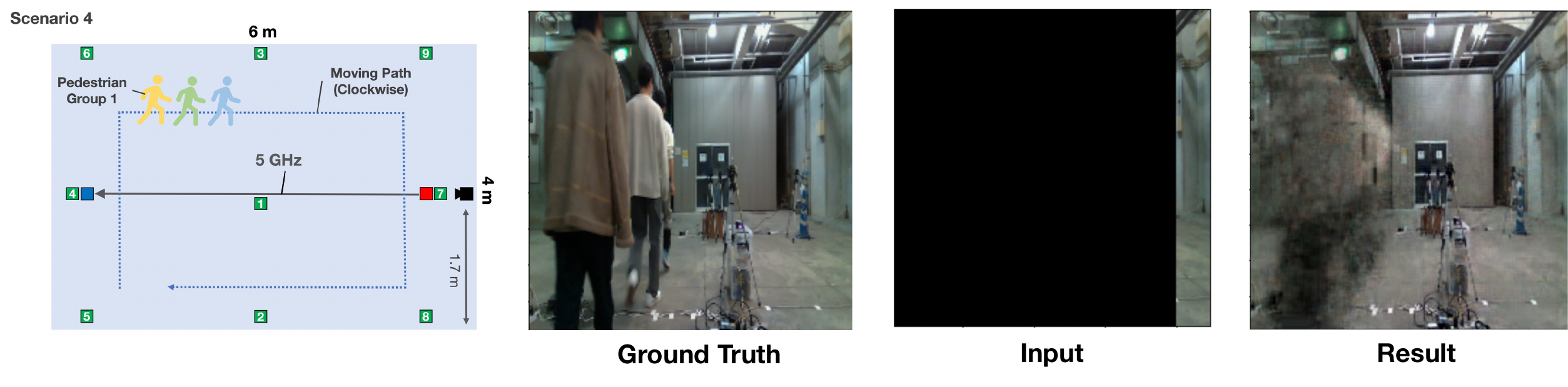}
        \subcaption{Result of Scenario 4.}
        \label{fig:factory_result_04}
    \end{minipage}
    \\ 
    
    \begin{minipage}{\linewidth}
        \centering
        \includegraphics[scale=0.55]{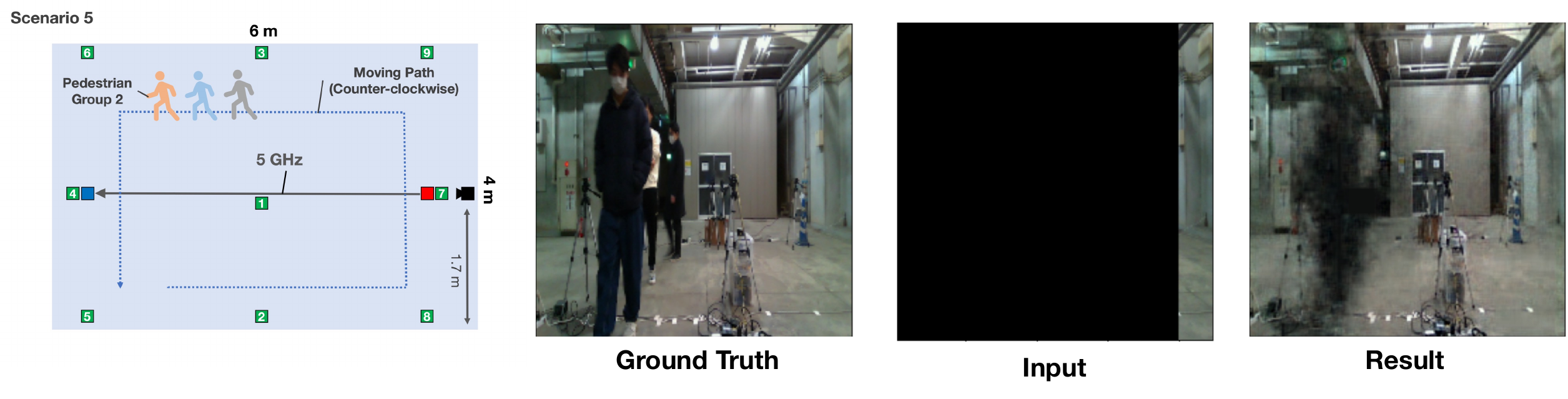}
        \subcaption{Result of Scenario 5.}
        \label{fig:factory_result_05}
    \end{minipage}
    \\ 

    \begin{minipage}{\linewidth}
        \centering
        \includegraphics[scale=0.55]{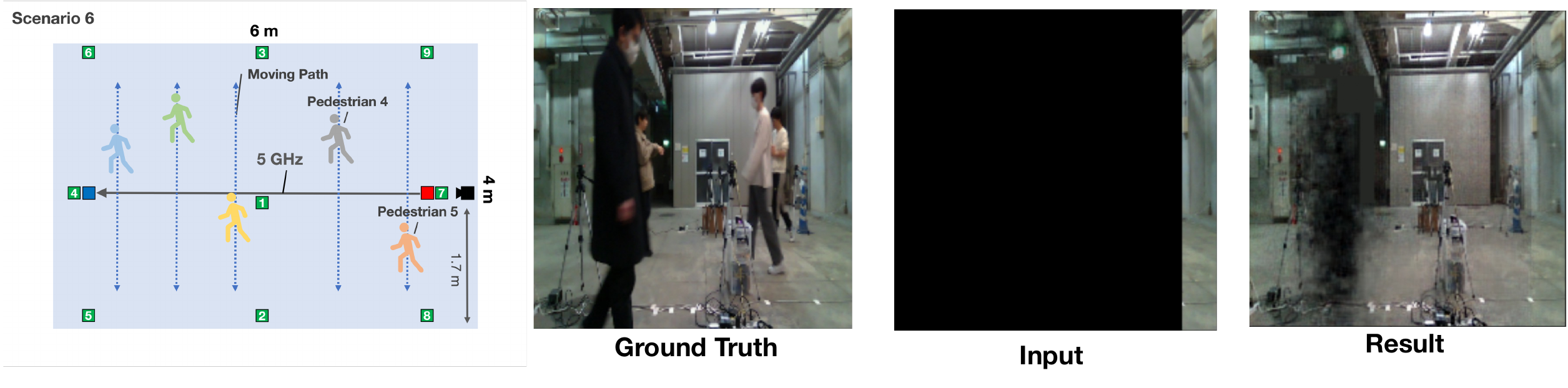}
        \subcaption{Result of Scenario 6.}
        \label{fig:factory_result_06}
    \end{minipage}
    \caption{
    \label{fig:Multiple_Pedestrian_results} Obstacle Removal for Multiple-Pedestrian Scenarios.}
\end{figure*}

The CSI-Inpainter's robustness was put to the test in a complex factory workshop setting to evaluate its adaptability to changing environments. The model initially trained in the office environment underwent fine-tuning across different scenarios, demonstrating its capability to adjust and maintain performance in more dynamic settings.

Fine-tuning commenced with 30 epochs on a single-person scenario (S1), necessitating more time due to the significant environmental change from the initial training setup. Subsequent scenarios involving different pedestrian movements (S2 and S3) required fewer epochs (10 each), suggesting the model's increasing efficiency in adapting to the workshop's conditions. Visual results show precise recovery of pedestrian features, indicating the model's high fidelity in replicating real-world details (\cref{fig:Single_Pedestrian_results}).

Post-fine-tuning, the model was re-evaluated on earlier scenarios (S1 and S2) to determine its ability to retain pre-trained information. The results were promising, showing accurate reconstruction of pedestrian appearance and position, despite slight color inconsistencies in scenario S2 (\cref{fig:revisited_scenarios}).

The investigation was expanded to encompass scenarios with multiple pedestrians (S4-S6), continuously recording data to challenge the CSI-Inpainter with increased complexity. In scenarios with three (S4, S5) and five pedestrians (S6), the model successfully reconstructed the color and location for all subjects (\cref{fig:Multiple_Pedestrian_results}). The use of multiple CSI sensors was pivotal in accurately rendering obstructed individuals, an achievement that traditional camera-based methods cannot claim. However, scenario S6 presented unique challenges, as the model favored imaging closer subjects over those positioned further away.

This extended evaluation in a factory workshop underscores the CSI-Inpainter's potential for wide-ranging applications, particularly in industrial settings where visual obstructions are common. Future research may explore enhancements such as increasing CSI resolution and incorporating depth information to overcome the current limitations identified in scenario S6.

\section{Impact of CSI Changes on Imaging Performance}

The efficacy of CSI-Inpainter in removing obstacles is intricately linked to the precision of CSI-driven imaging. Our exploration into CSI's imaging capabilities is guided by the hypothesis that variations in CSI data significantly influence the quality of obstacle removal. To substantiate this, we conducted a series of experiments, manipulating CSI data inputs in terms of sensor location, data fusion from multiple sensors, and modifications across temporal and frequency dimensions of the CSI matrix. These experiments aim to deepen our understanding of how CSI data intricacies affect the imaging process, which is central to the obstacle removal performance of CSI-Inpainter.

\subsection{Imaging with CSI from Individual Sensors at Different Locations}

\begin{figure}[h]
    \begin{tabular}{cc}
    
      \begin{minipage}[h]{\linewidth}
        \centering
        \includegraphics[scale=0.35]{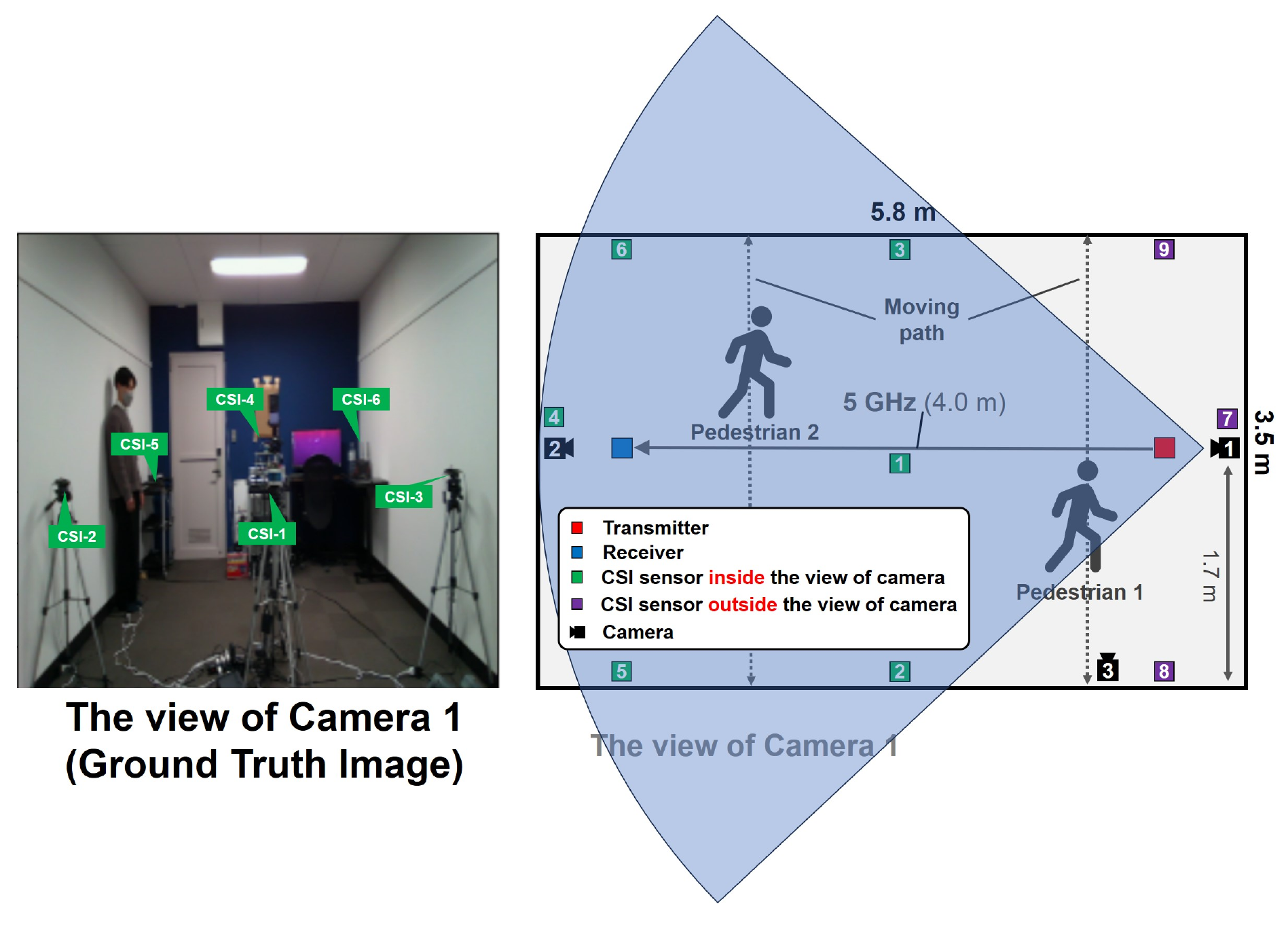}
        \subcaption{Camera 1 View: captured CSI sensors vs. non-captured sensors. Only Group 3 is excluded from the view.}
        \label{fig:c1_gd}
      \end{minipage}\\

      \begin{minipage}[h]{\linewidth}
        \centering
        \includegraphics[scale=0.36]{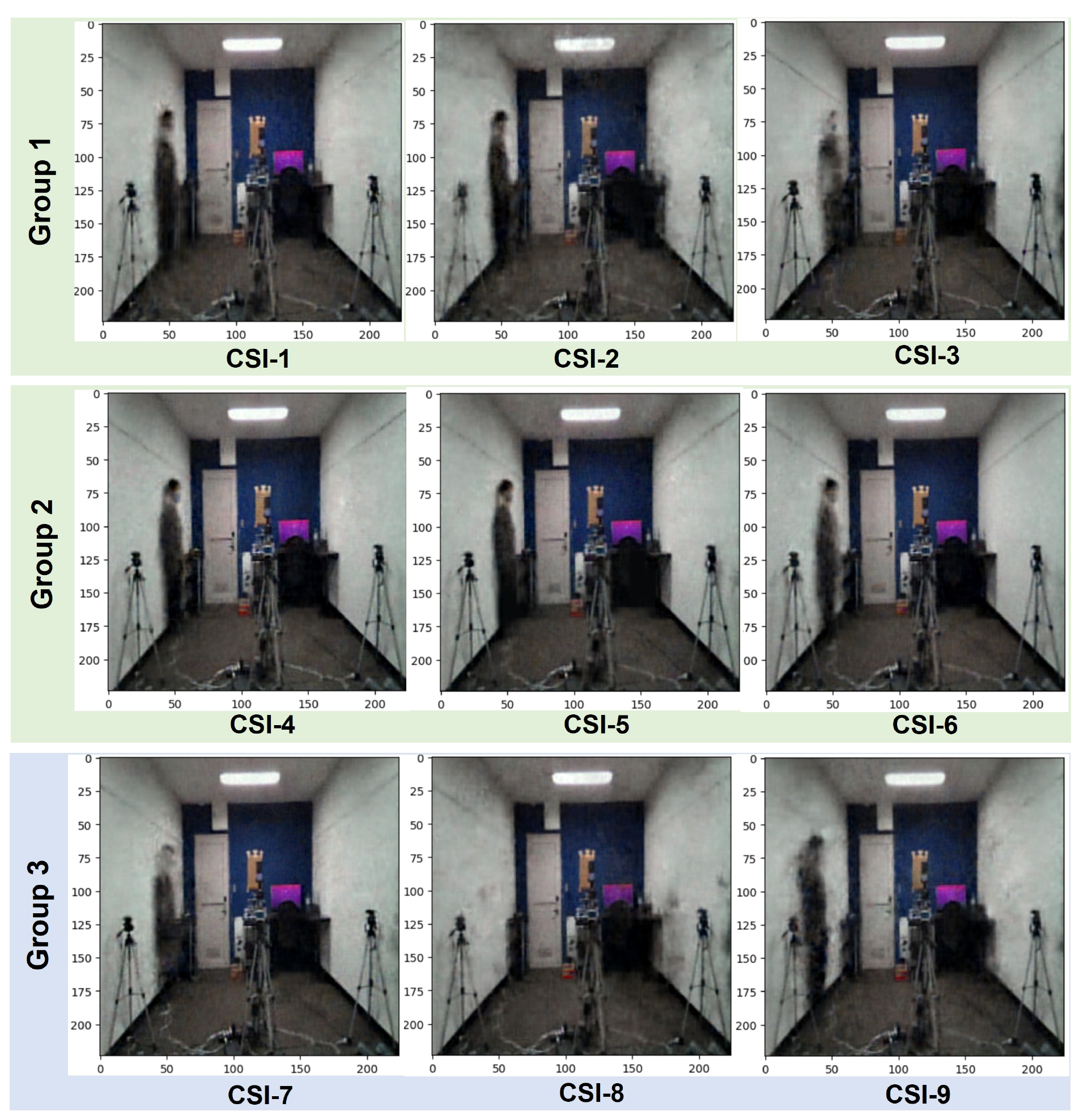}
        \subcaption{The RF-imaging result of each sensor.}
        \label{fig:c1_img}
      \end{minipage}\\
    \end{tabular}
    \caption{Variation in RF-only imaging results for Camera 1 using single-sensor CSI at different locations. Visually, Group 2 surpasses Group 1, while Group 1 outperforms Group 3.}
    \label{fig:c1_impact} 
  \end{figure}

\begin{figure}[h]
    \begin{tabular}{cc}
    
      \begin{minipage}[h]{\linewidth}
        \centering
        \includegraphics[scale=0.36]{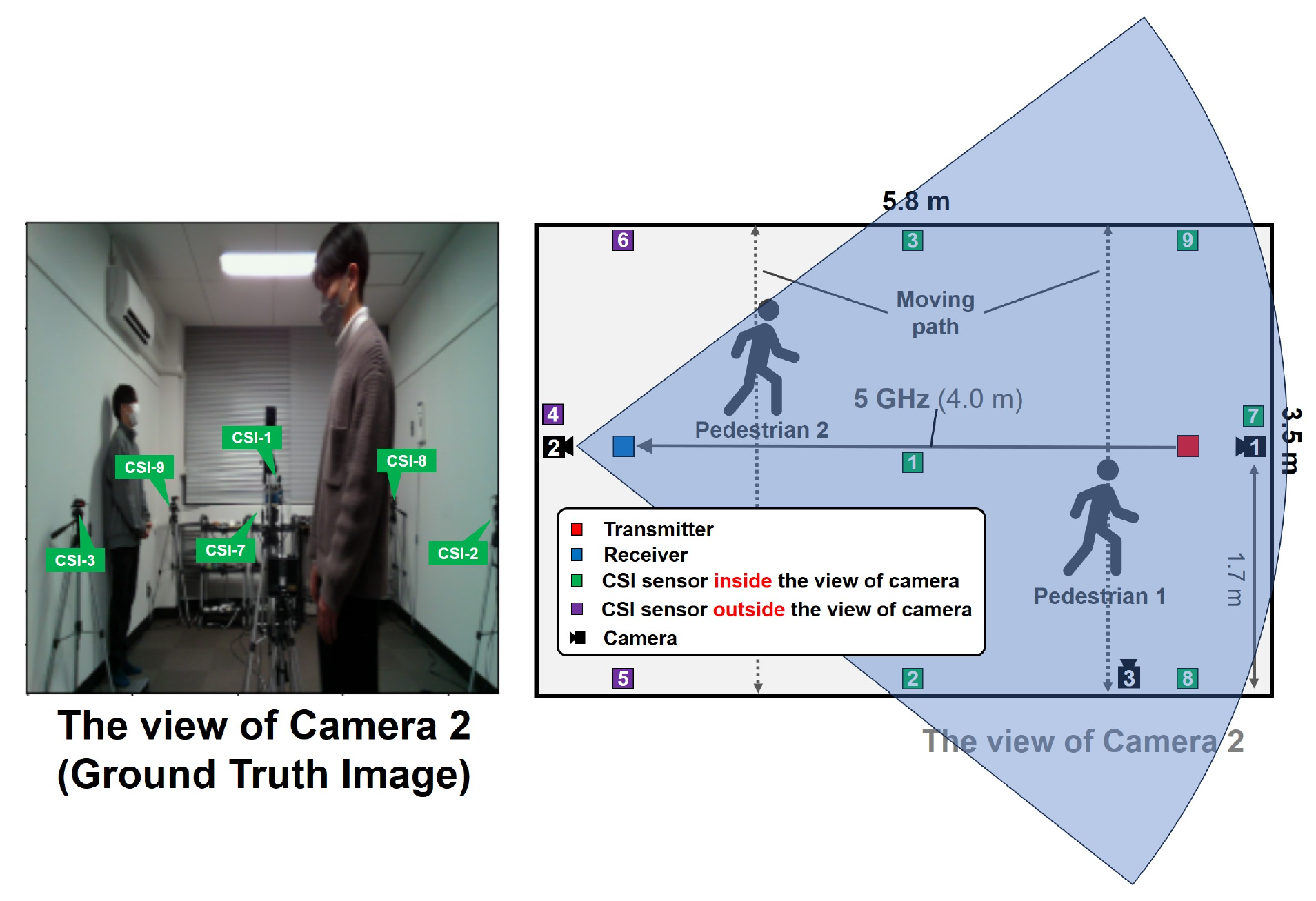}
        \subcaption{Camera 2 View: captured CSI sensors vs. non-captured sensors. Only Group 2 is excluded from the view.}
        \label{fig:c2_gd}
      \end{minipage}\\

      \begin{minipage}[h]{\linewidth}
        \centering
        \includegraphics[scale=0.36]{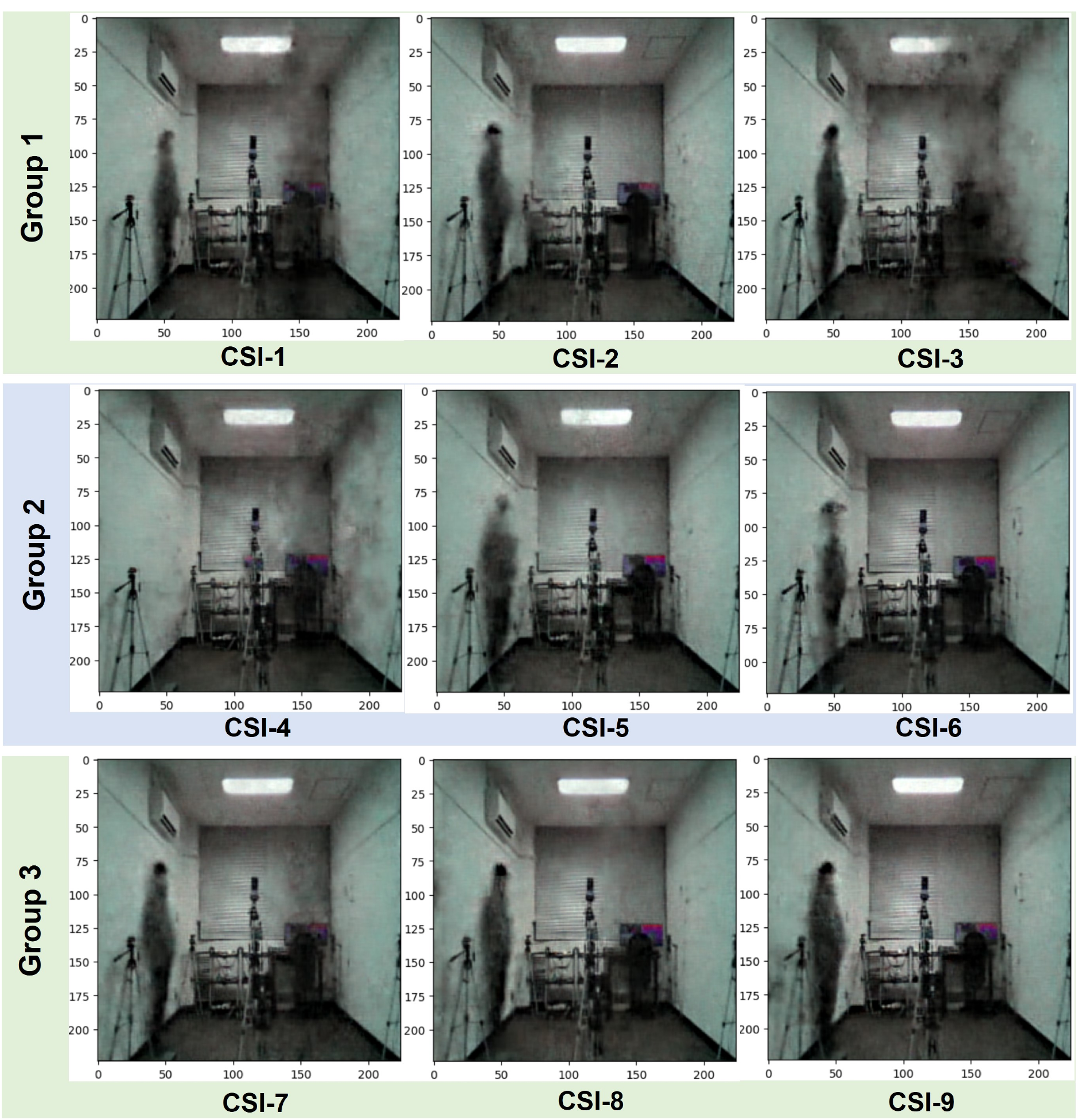}
        \subcaption{The RF-imaging result of each sensor.}
        \label{fig:c2_img}
      \end{minipage}\\

    \end{tabular}
    \caption{Variation in RF-only imaging results for Camera 2 using single-sensor CSI at different locations. Visually, Group 3 surpasses Group 1, while Group 1 outperforms Group 2.}
    \label{fig:c2_impact} 
\end{figure}

\begin{figure}[t]
    \begin{tabular}{cc}
    
      \begin{minipage}{\linewidth}
        \centering
        \includegraphics[scale=0.4]{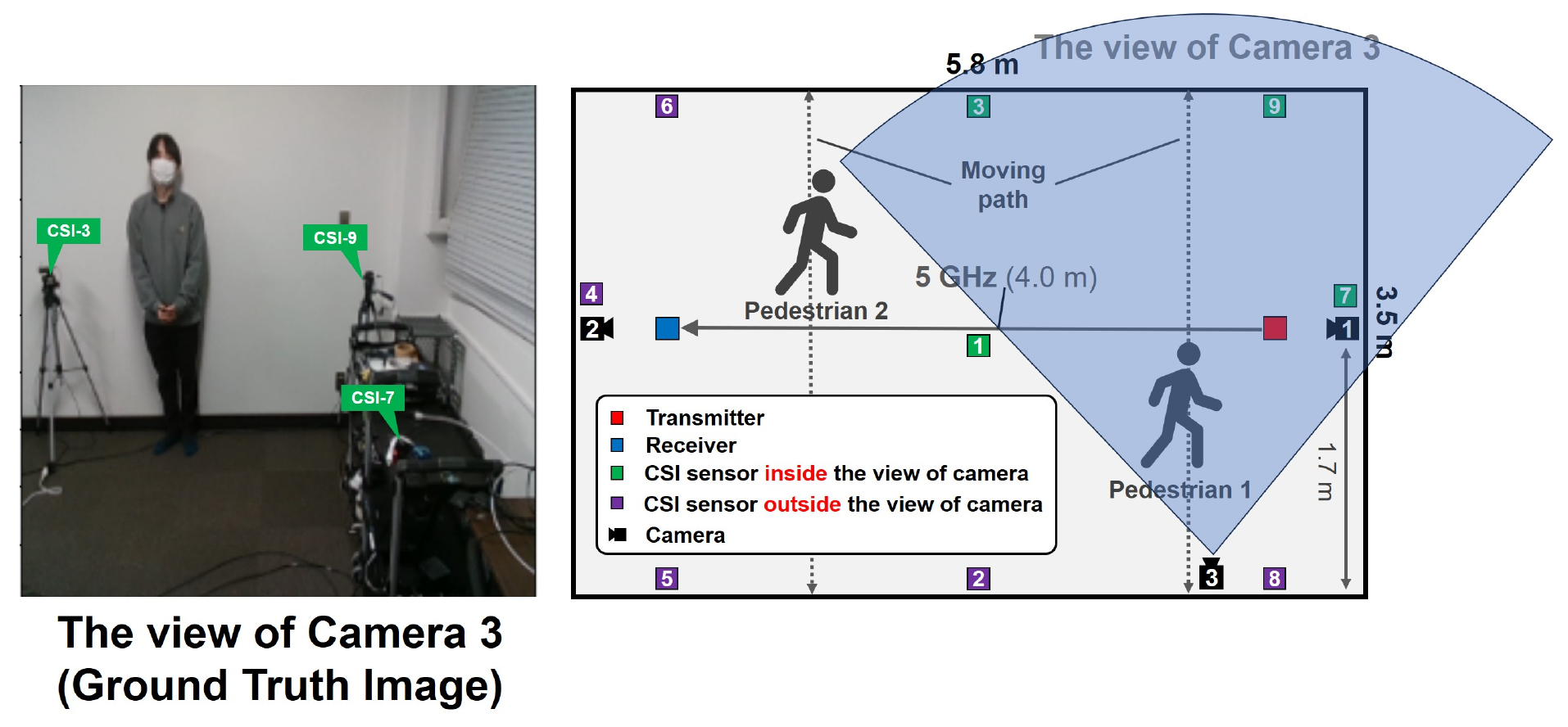}
        \subcaption{Camera 3 View: captured CSI sensors vs. non-captured sensors. CSI-1, CSI-3, CSI-7, and CSI-9 are included in the view.}
        \label{fig:c3_gd}
      \end{minipage}\\

      \begin{minipage}{\linewidth}
        \centering
        \includegraphics[scale=0.4]{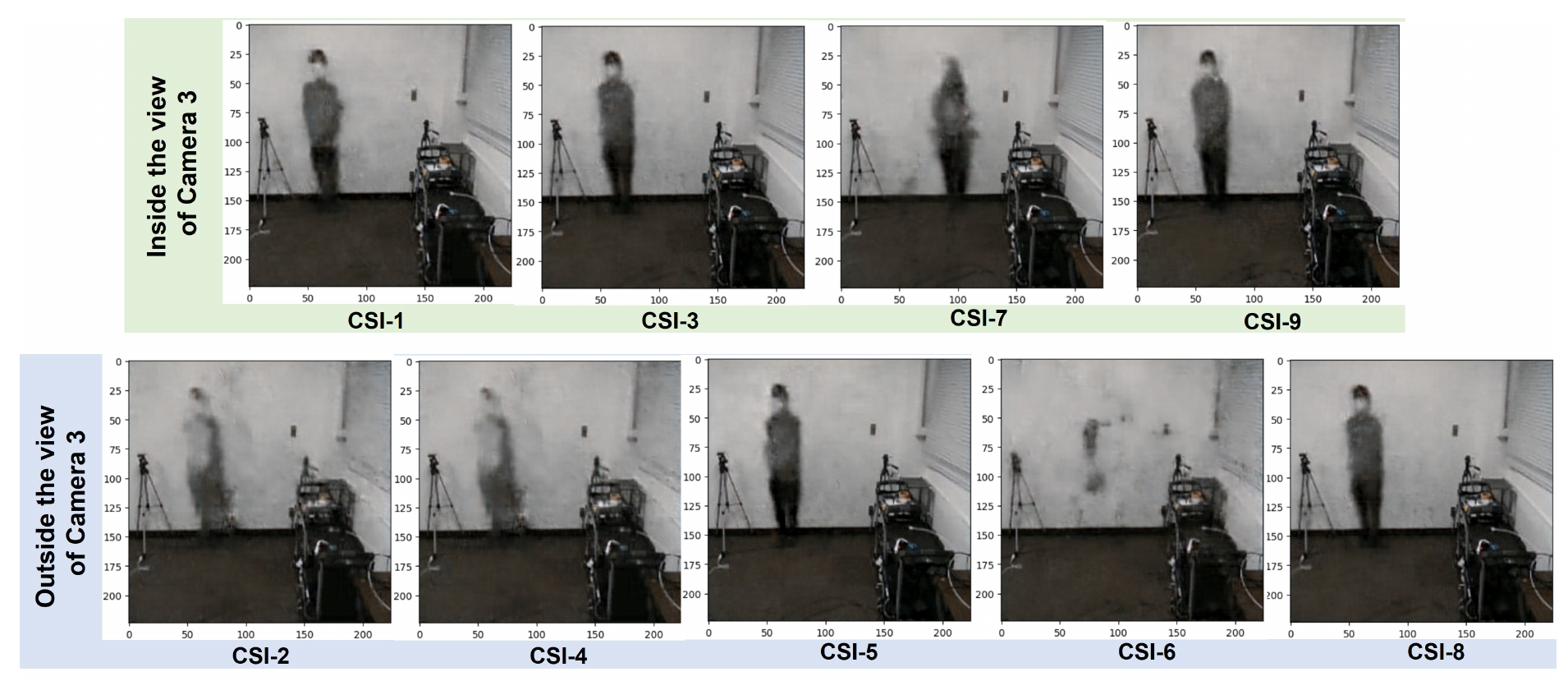}
        \subcaption{The RF-imaging result of each sensor.}
        \label{fig:c3_img}
      \end{minipage}\\

    \end{tabular}
    \caption{Variation in RF-only imaging results for Camera 3 using single-sensor CSI at different locations. Generally, harnessing CSI from sensors within the field of view of Camera 3 can enhance imagery quality.}
    \label{fig:c3_impact} 
\end{figure}

\begin{figure*}[t]
    \centering
    \includegraphics[scale=0.8]{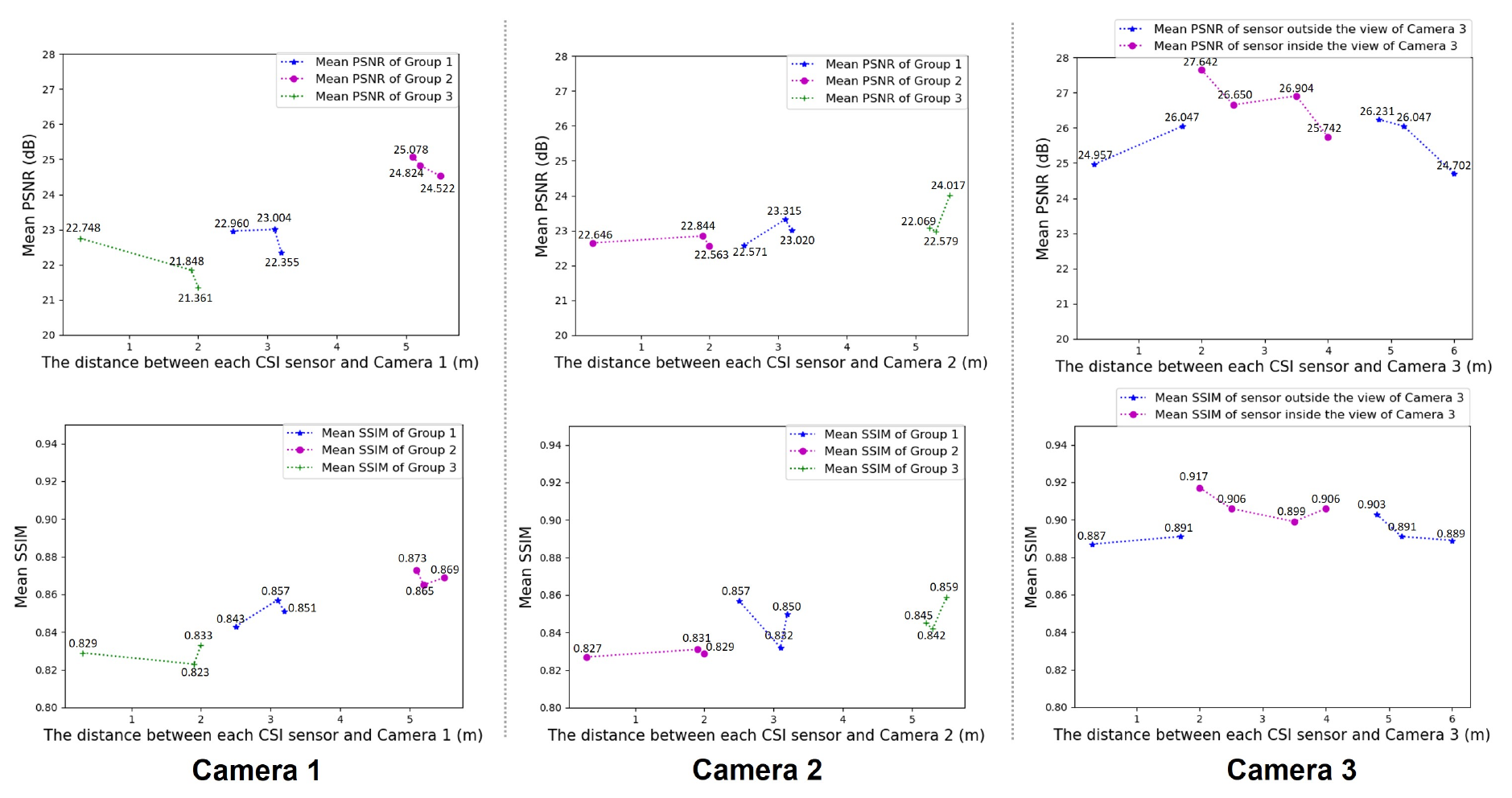}
    \caption{
    Quantitative evaluation of CSI-Inpainter's imaging performance across three camera datasets at varying CSI sensor-to-camera distances. Generally, imaging within CSI captured within the camera's field yields enhanced PSNR and SSIM metrics. Furthermore, a marginal increase in sensor-to-camera distance correlates with further improved scores.}
    \label{fig:obj} 
\end{figure*}

In the pursuit of advancing CSI-based imaging for obstacle removal, our investigation meticulously examines the impact of CSI sensor placement within the office setting. This locale is selected to mitigate external noise and interference, thus ensuring the integrity of the CSI data.

We organized the CSI sensors into strategic groupings relative to Cameras 1 and 2, denoted as Group 1 (CSI-1, CSI-2, CSI-3), Group 2 (CSI-4, CSI-5, CSI-6), and Group 3 (CSI-7, CSI-8, CSI-9). As delineated in Figures \cref{fig:c1_gd} and \cref{fig:c2_gd}, Groups 1 and 2 are positioned within Camera 1's observational scope, whereas Group 3 resides beyond it. For Camera 2, visibility is obstructed only for Group 2.

\subsubsection{Camera 1}

The results of the Camera 1 dataset are shown in \cref{fig:c1_img} and \cref{fig:obj}. Overall, we obtained better imaging results, both visually and objectively, by using CSI from sensors inside the view of Camera 1 rather than outside. Groups 1 and 2 show significantly better imaging quality than Group 3, as evidenced by higher mean PSNR and SSIM scores. Additionally, as the distance between the CSI sensor and the camera increases (from $0.3 \text{m}$ to $6.5 \text{m}$), the imaging results improve. The mean PSNR and SSIM scores of Group 2 reach their peak ($25.078 \text{dB}$ in PSNR and $87.3 \%$ in SSIM), corresponding to the clearest images in \cref{fig:c1_img}. Group 1 results in a slight degradation of imaging quality compared to Group 2 (approximately $2 \text{dB}$ in PSNR and $2 \%$ in SSIM). Conversely, Group 3 exhibits the lowest scores ($21.361 \text{dB}$ in PSNR and $82.3 \%$ in SSIM) in \cref{fig:obj}, corresponding to the worst imaging quality in \cref{fig:c1_img}.

The rationale behind these observations is that the CSI data collected within the camera's field of view more adequately captures the spatial, physical information within that localized environment, enhancing the imaging quality. Moreover, appropriately increasing the distance between the CSI sensor and the camera within a certain range allows the sensor to collect more Wi-Fi signals reflected back from different angles within the room, providing a more comprehensive perception of the environment.

However, it is important to note that the distance between the CSI sensor and the camera should not be too far, even if the sensor is located within the camera's field of view, as the intensity of radio signals along the propagation path may vary. Longer propagation paths introduce more noise and interference into the wireless data, leading to significant degradation of imaging accuracy. Therefore, determining the optimal position relation between CSI sensors and the camera for indoor RF-based imaging warrants further investigation in future studies.

\subsubsection{Camera 2}
The experimental results of Camera 2 align with those of Camera 1, given the similarity in the position relation between Camera 2 and the CSI sensors. Specifically, Group 3 achieves the highest imaging performance ($24.017 \text{dB}$ in PSNR and $85.7 \%$ in SSIM) as the sensors in this group are not only within the field of view of Camera 2 but also located at the farthest distance ($6.5 \text{m}$ or so) from the camera. As the sensors move closer to the camera (from $6.5 \text{m}$ to $0.3 \text{m}$) and then outside its field of view, the imaging quality gradually degrades ($1.454 \text{dB}$ in PSNR and $3.2 \%$ in SSIM), as shown in \cref{fig:c2_img} and \cref{fig:obj}.

Furthermore, a comparison of the results in \cref{fig:obj} reveals that the imaging quality of Camera 2 is slightly lower than that of Camera 1 ($1.061 \text{dB}$ in PSNR and $1.4 \%$ in SSIM). This difference may be attributed to the presence of two pedestrians captured by Camera 2. With visual information from a single CSI sensor, it becomes extremely challenging to accurately reconstruct the figures of two persons distributed at different locations in the room. In contrast, Camera 1 focuses on only one moving pedestrian, making the imaging task relatively easier to handle.

\subsubsection{Camera 3}
The imaging view of Camera 3 is primarily covered by the sensing ranges of CSI-1, CSI-3, CSI-7, and CSI-9 (although CSI-1 is not shown in the image, its positional relation to Camera 3 is similar to that of CSI-7). The results indicate that the best imaging performances ($27.642 \text{dB}$ in PSNR and $91.7 \%$ in SSIM) are more likely to be obtained with the CSI from these sensors, as shown in \cref{fig:c3_impact} and \cref{fig:obj}. Using the CSI data collected from sensors outside the camera's view results in a maximal degradation of image quality, with a decrease of about $2.94 \text{dB}$ in PSNR and $3.0 \%$ in SSIM.

Moreover, by calculating the average PSNR and SSIM for the nine CSI-guided imaging results on each camera dataset, we observe that Camera 3 (PSNR: $25.885 \text{dB}$, SSIM: $89.9 \%$) achieves significantly better results than Camera 1 (PSNR: $23.189 \text{dB}$, SSIM: $84.9 \%$) and Camera 2 (PSNR: $23.002 \text{dB}$, SSIM: $84.2 \%$). This improvement may be because Camera 3 focuses on less than $1/4$ of the room's space (a corner, actually), while Camera 1 and Camera 2 cover over $2/3$ of the room. As a result, Camera 3 requires less imagery information from CSI to visualize the environment.

\subsection{Variation of CSI Matrix}

The CSI matrix, representing the channel state over different subcarriers and time slots, undergoes variations in the time or frequency dimension, significantly impacting imaging performance.

\subsubsection{Changes in the Time Dimension}

\begin{figure}[t]
    \centering
    \includegraphics[scale=0.5]{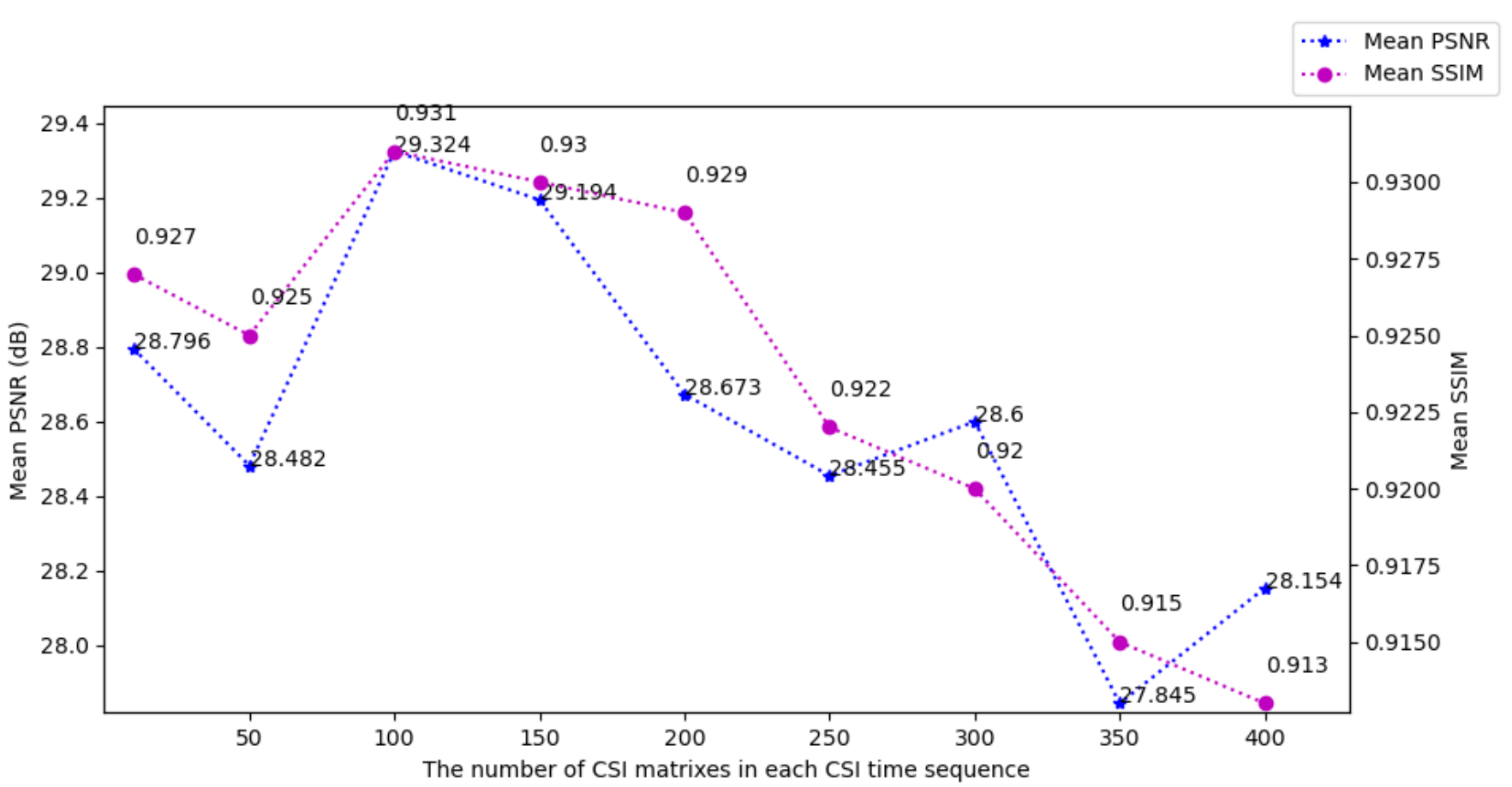}
    \caption{\label{fig:matrix-num} Progressive enhancement and decline of imaging quality in CSI-Inpainter with increasing time window length.}
\end{figure}

In our previous work \cite{chen2022rf}, we demonstrated the significance of selecting an appropriate number of RSSI values in a single time sequence to achieve effective image inpainting. Here, we explore the impact of varying the length ($L$) of the time sequence of the CSI matrix (ranging from $L=10$ to $L=400$ with an interval of 50) obtained from CSI-1, CSI-3, CSI-7, and CSI-9 for reconstructing Camera 3 images. The imaging results in \cref{fig:matrix-num} show a gradual improvement as the time window length increases, reaching its peak at around $L=150$, followed by a gradual decline. This observation aligns with our findings in \cite{chen2022rf}. Surprisingly, clear imaging results can even be obtained with a CSI time dimension of 10 ($L=10$), an achievement impossible with RSSI, reaffirming the superiority of CSI over RSSI in this context.

\subsubsection{Subcarrier-dimension Compression}

\begin{figure}[t]
    \centering
    \includegraphics[scale=0.4]{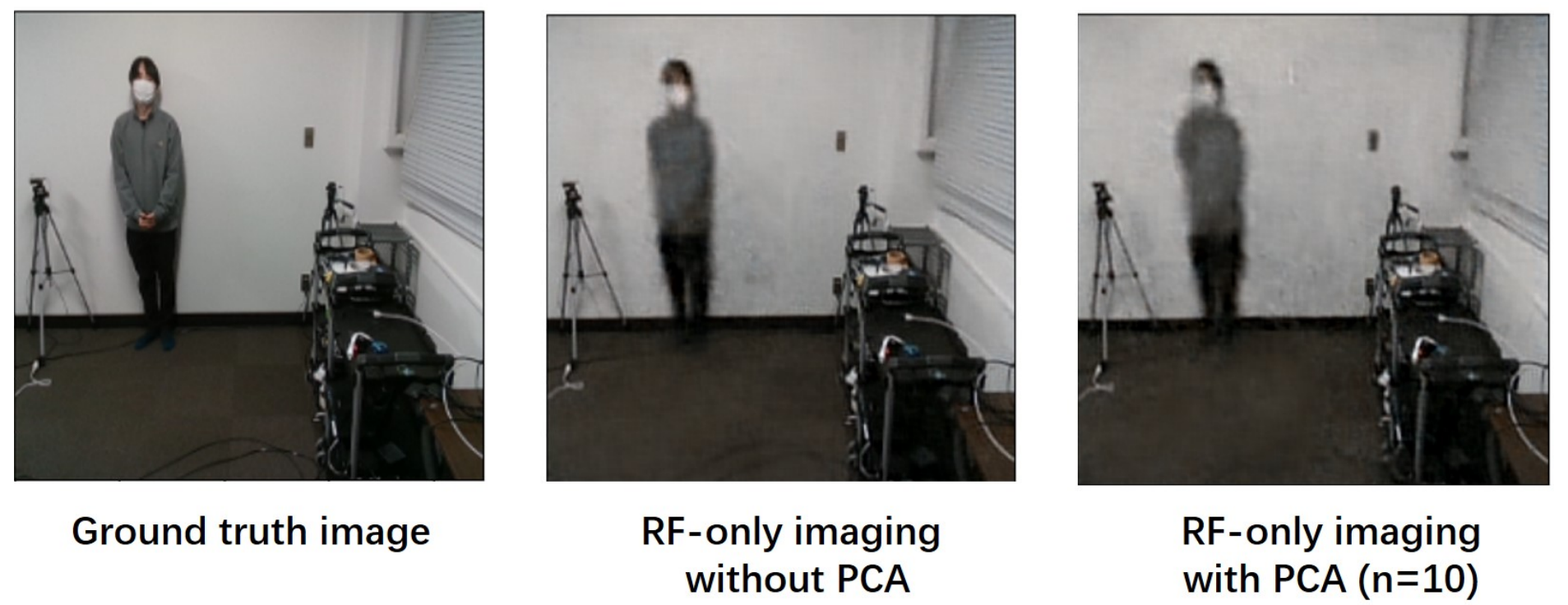}
    \caption{Effect of PCA on CSI for CSI-Inpainter: minimal impact on imaging quality.}
    \label{fig:combination} 
\end{figure}

The frequency dimension indicates the number of subcarriers a Wi-Fi channel with MIMO can be divided into using OFDM. While a higher number of subcarriers theoretically provides richer CSI information and better imaging performance, it also increases computational resources and time consumption, especially when imaging with CSI matrices from multiple sensors.

To achieve an effective and computationally friendly imaging task, we consider finding a balance between imaging performance and the frequency dimension of the CSI matrices. We perform PCA on the frequency dimension of CSI matrices. PCA is a statistical technique used to reduce the dimensionality of a dataset by identifying its most essential features or components. In the context of CSI matrices, PCA can reduce the frequency dimension while retaining essential information about the wireless channel's characteristics.

Our experimental results demonstrate the feasibility of this approach. Using CSI matrices from CSI-1, CSI-3, CSI-7, and CSI-9 to image from the viewpoint of Camera 3, we applied PCA to reduce the frequency dimension from 256 to 10, leading to a compression of hyper-parameters in CSI-Inpainter from 434,270,379 to 23,588,529 (a compression of $94.6\%$). Consequently, the training time of CSI-Inpainter decreased from $1974.395 \text{s}$ to $951.483 \text{s}$ (a compression of $51.8 \%$). Visually, as shown in \cref{fig:combination}, the imaging results were not significantly affected, with only a slight decrease in mean SSIM (from $92.955 \%$ to $91.870 \%$) and mean PSNR (from $29.040 \text{dB}$ to $28.553 \text{dB}$). This trade-off between imaging performance and computation time proves worthwhile, particularly when dealing with a substantial amount of training data.

\subsection{Fusing CSI from Multiple Sensors}

\begin{figure}[t]
\begin{tabular}{cc}
  \begin{minipage}{\linewidth}
    \centering
    \includegraphics[scale=0.45]{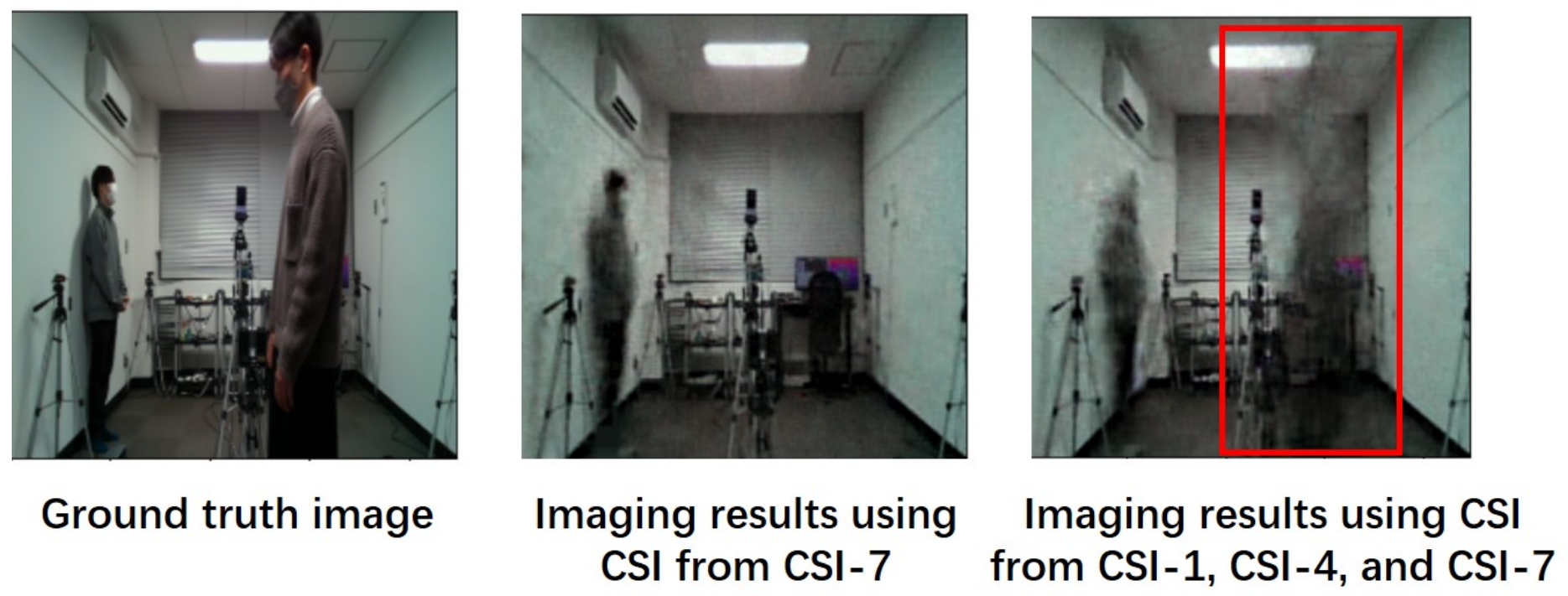}
    \subcaption{Enhanced pedestrian portrait reconstruction using integrated CSI data from multiple sources.}
    \label{fig:office_combination}
  \end{minipage}\\
  
  \begin{minipage}{\linewidth}
    \centering
    \includegraphics[scale=0.45]{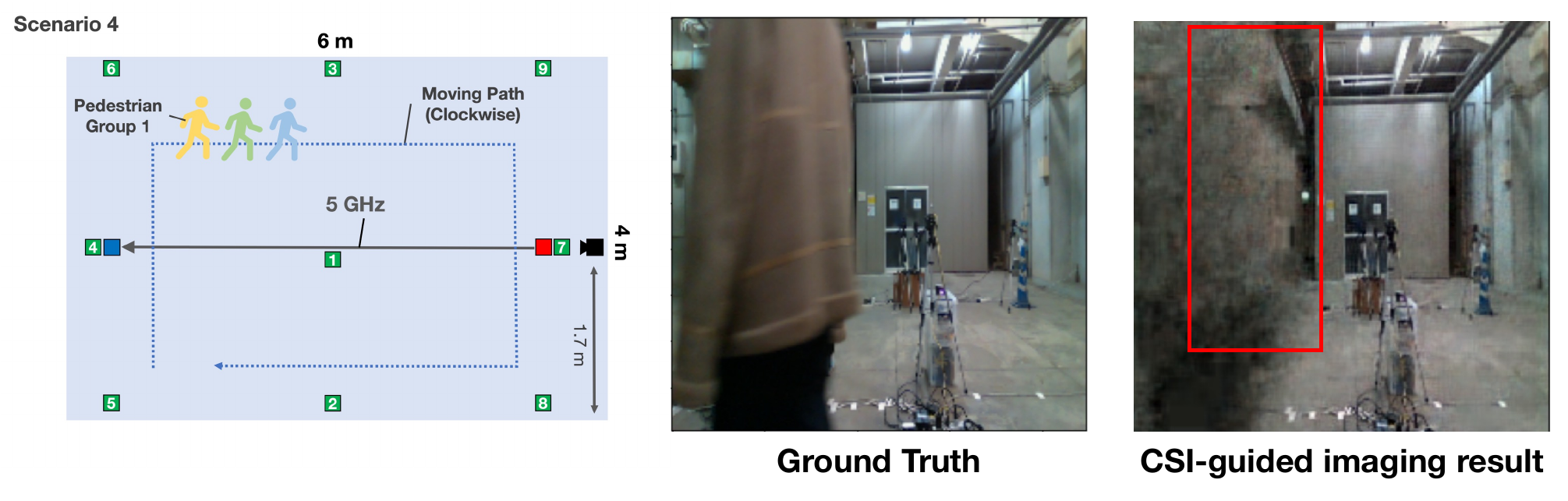}
    \subcaption{
    \label{fig:fac_double_inpainting_1} Obstructed Imaging Recovery 1: Detailed imaging of concealed figures through multi-sensor CSI fusion.}
    
  \end{minipage}\\

  \begin{minipage}{\linewidth}
    \centering
    \includegraphics[scale=0.45]{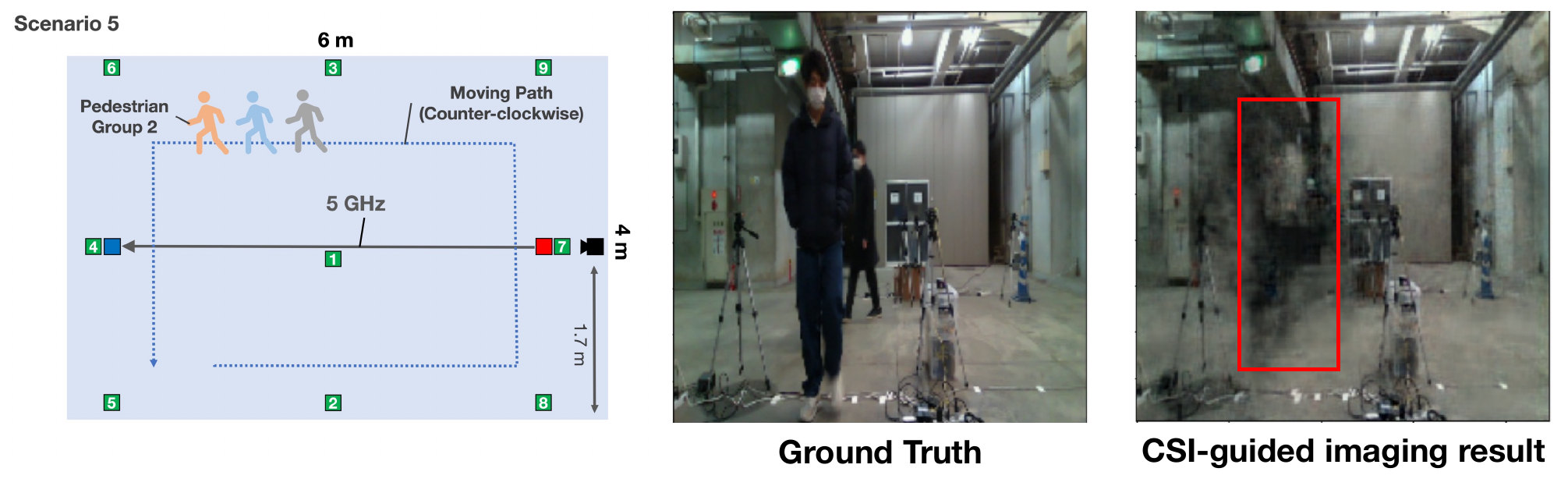}
    \subcaption{
    \label{fig:fac_double_inpainting_2} Obstructed Imaging Recovery 2: Detailed imaging of concealed figures through multi-sensor CSI fusion.}
  \end{minipage}\\
\end{tabular}
\caption{\label{fig:fac_double_inpainting} Comprehensive Obstacle Imaging with Multisensor CSI Integration}
\end{figure}

The preceding subsection demonstrated that while a single sensor's CSI can yield precise imaging results when the sensor is near the imaging area, it has limitations in achieving a comprehensive perception and imaging of the entire environment.

To illustrate this issue, we refer to the experimental results shown in \cref{fig:office_combination}. From the viewpoint of Camera 2, two pedestrians can be observed moving at different distances, with one closer and the other farther away. Initially, we employ only the CSI data from CSI-7 for imaging, which is closer to the distant pedestrian ($0.5 \text{m}$ or so). In this case, CSI-Inpainter successfully recovers the image of the distant pedestrian but fails to reconstruct the image of the closer pedestrian. However, by incorporating additional CSI data from CSI-1 and CSI-4, we resolve this problem, leading to a precise reconstruction of both pedestrians' portraits. This observation leads us to conclude that the final imaging range results from the combination of the perceptual ranges of individual sensors. To achieve a complete reconstruction of an image taken by a camera, it is essential to utilize multiple sensors to ensure that the perceptual range of RF signals fully covers the entire camera view.

Furthermore, this conclusion explains the consistently better results obtained from Camera 3 in Section 4 compared to Cameras 1 and 2. Generally, Camera 3 captures images of only one person, while the other two capture images of two people. When using the CSI from a single sensor for imaging, it may lead to losing some portrait information, resulting in lower Mean PSNR and SSIM scores. This difference in imaging outcomes highlights the importance of using multiple sensors to achieve comprehensive and accurate image reconstruction.

Additional experiments, as depicted in \cref{fig:fac_double_inpainting}, supplement our findings from the factory environment. Here, we explore the imaging capability of CSI-Inpainter when confronted with obstructed figures. \cref{fig:fac_double_inpainting_1} demonstrates that, despite obstructions from P3, the fusion of CSI data from sensors CSI-1, 2, 3, 4 enables a partial recovery of obstructed P1 and P2, illustrating the model's capability to reconstruct figures even when they are not fully visible. Similarly, \cref{fig:fac_double_inpainting_2} shows that by combining CSI data from the same sensor array, the model can reconstruct the figure of P1, which is obscured by P5. These outcomes validate the potential of CSI-Inpainter in complex multi-person scenarios and underscore the importance of leveraging data from multiple CSI sensors to ensure comprehensive and accurate obstacle removal in varied environments. 

Besides, this discovery inspires an area for further research, where enhancing CSI density and leveraging depth information (depth cameras) may provide solutions. Incorporating depth information into CSI data analysis for obstacle removal could significantly enhance imaging accuracy by providing a three-dimensional understanding of the environment, thus allowing for precise distinction and reconstruction of obscured or overlapping objects in complex scenarios.

\section{Conclusion}

This research presents CSI-Inpainter, a novel approach to obstacle removal leveraging CSI time sequences and multimodal Transformer architecture. Through extensive experimentation in both office and factory environments, we demonstrated the superior capability of CSI-Inpainter to accurately reconstruct obstructed images, significantly outperforming traditional methods and our previous work, RF-Inpainter, in terms of visual quality and objective metrics such as PSNR and SSIM.

Key findings include the advantage of using CSI data over RSSI for richer visual information, the critical role of sensor placement and data fusion from multiple CSI sensors in enhancing imaging outcomes, and the adaptability of CSI-Inpainter to diverse and complex scenarios. Notably, the integration of CSI from multiple sensors was found to be crucial for comprehensive perception and imaging, especially in scenarios with multiple occlusions.

Although this work presents promising results, several avenues for future research can be explored. Firstly, adopting advanced Transformer variants and architecture modifications may further enhance CSI-Inpainter's performance. Additionally, exploring different fusion strategies to combine visual and temporal data and integrating other sensor modalities, such as LiDAR or audio information, could lead to even more robust and versatile obstacle removal approaches. Furthermore, extending this research to real-world applications with diverse and complex environments, such as outdoor scenarios, would be of significant interest. Evaluating CSI-Inpainter's performance in other wireless sensing tasks, such as motion tracking and localization, could also uncover its potential for broader applications in wireless communication and CV.

\section{Acknowledgments}

This work was supported in part by JSPS KAKENHI Grant Number JP22H03575.

\ifCLASSOPTIONcaptionsoff
  \newpage
\fi
\bibliographystyle{IEEEtran.bst}
\bibliography{main_text.bib}

\clearpage

\begin{IEEEbiography}[{\includegraphics[width=1.1in,height=1.3in, clip,keepaspectratio]{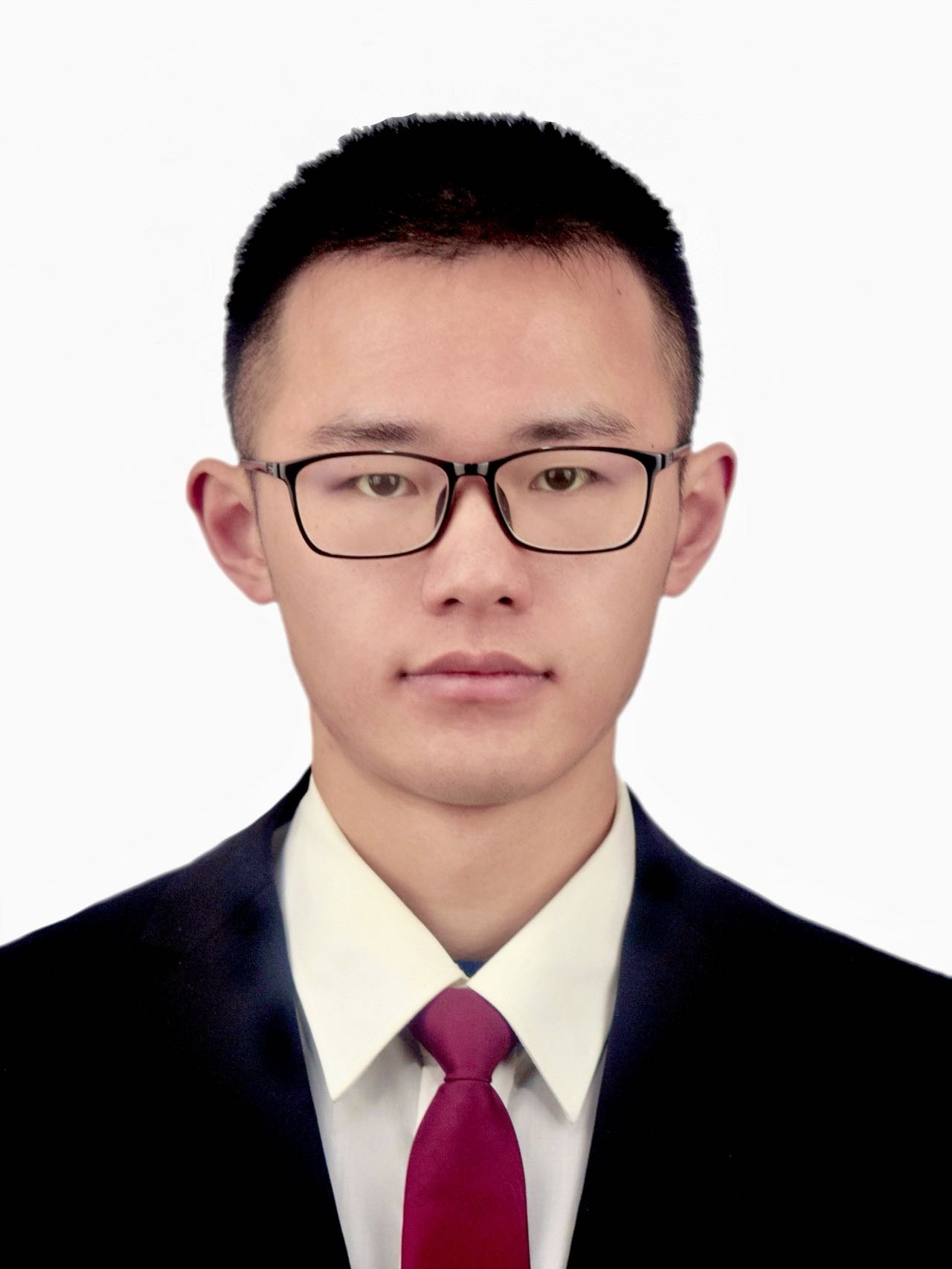}}]{Cheng Chen} received the B.E. degree in Information and Communications Engineering (ICE) from the National University of Defence Technology in 2019 and the master’s degree in ICE from Tokyo Institute of Technology in 2023. He received the Outstanding Student Award from the Department of ICE in 2023 and is currently undertaking a Ph.D. at Tokyo Institute of Technology. 
\end{IEEEbiography}

\begin{IEEEbiography}[{\includegraphics[width=1.1in,height=1.3in, clip,keepaspectratio]{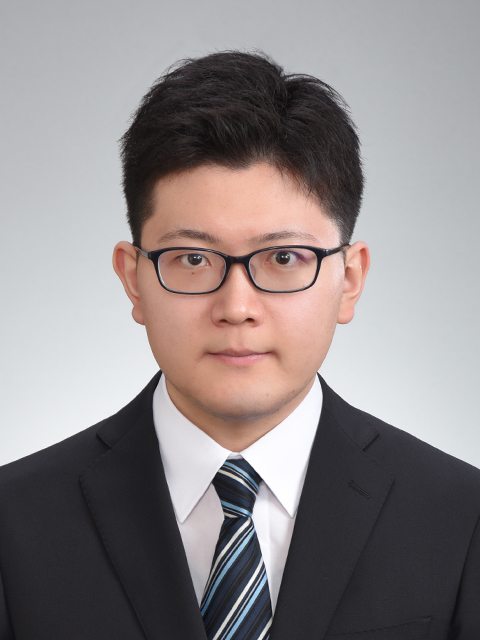}}]{Shoki Ohta} (Student Member, IEEE) received the B.E.\ degree in Information and Communications Engineering from Tokyo Institute of Technology in 2022.
He is currently studying toward the M.E.\ degree at the School of Engineering, Tokyo Institute of Technology. He received the IEEE Vehicular Technology Society (VTS) Japan Young Researcher’s Encouragement Award in 2022.
\end{IEEEbiography}

\begin{IEEEbiography}[{\includegraphics[width=1.1in,height=1.3in, clip,keepaspectratio]{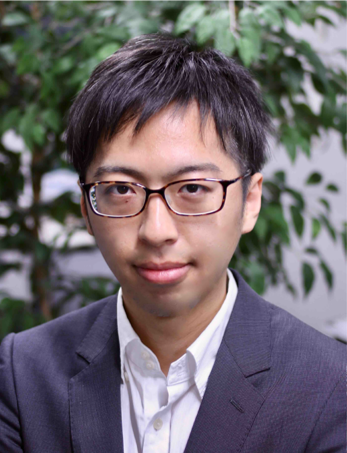}}]{Takayuki Nishio} (Senior Member, IEEE) received the B.E. degree in electrical and electronic engineering and the master's and Ph.D. degrees in informatics from Kyoto University, in 2010, 2012, and 2013, respectively. He was an Assistant Professor in communications and computer engineering at the Graduate School of Informatics, Kyoto University, from 2013 to 2020. From 2016 to 2017, he was a Visiting Researcher at the Wireless Information Network Laboratory (WINLAB), Rutgers University, USA. Since 2020, he has been an Associate Professor at the School of Engineering, Tokyo Institute of Technology, Japan, and the Wireless Information Network Laboratory (WINLAB), Rutgers University. His current research interests include machine learning-based network control, machine learning in wireless networks, vision-aided wireless communications, and heterogeneous resource management.
\end{IEEEbiography}


\begin{IEEEbiography}[{\includegraphics[width=1.1in,height=1.3in,clip,keepaspectratio]{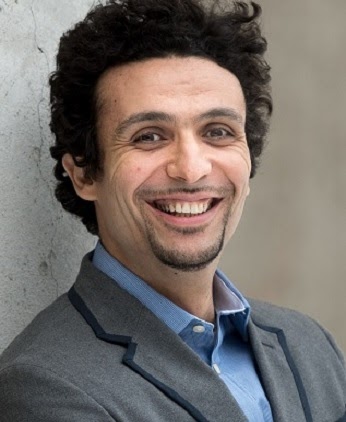}}]{Mehdi Bennis} (Fellow, IEEE) is currently a Professor with the Centre for Wireless Communications, University of Oulu, Finland, an Academy of Finland Research Fellow, and the Head of the Intelligent Connectivity and Networks/Systems Group (ICON). He has published over 200 research papers in international conferences, journals, and book chapters. His main research interests include radio-resource management, heterogeneous networks, game theory, and distributed machine learning in 5G networks and beyond. He has been a recipient of several prestigious awards, including the 2015 Fred W. Ellersick Prize from the IEEE Communications Society, the 2016 Best Tutorial Prize from the IEEE Communications Society, the 2017 EURASIP Best Paper Award for the Journal of Wireless Communications and Networking, the University of Oulu Award for Research, the 2019 IEEE ComSoc Radio Communications Committee Early Achievement Award, and the 2020 Clarviate Highly Cited Researcher from the Web of Science. He is also an editor of IEEE Transactions on Communications (TCOM) and the Specialty Chief Editor for Data Science for Communications in the Frontiers in Communications and Networks journal.
\end{IEEEbiography}

\begin{IEEEbiography}[{\includegraphics[width=1.1in,height=1.3in,clip,keepaspectratio]{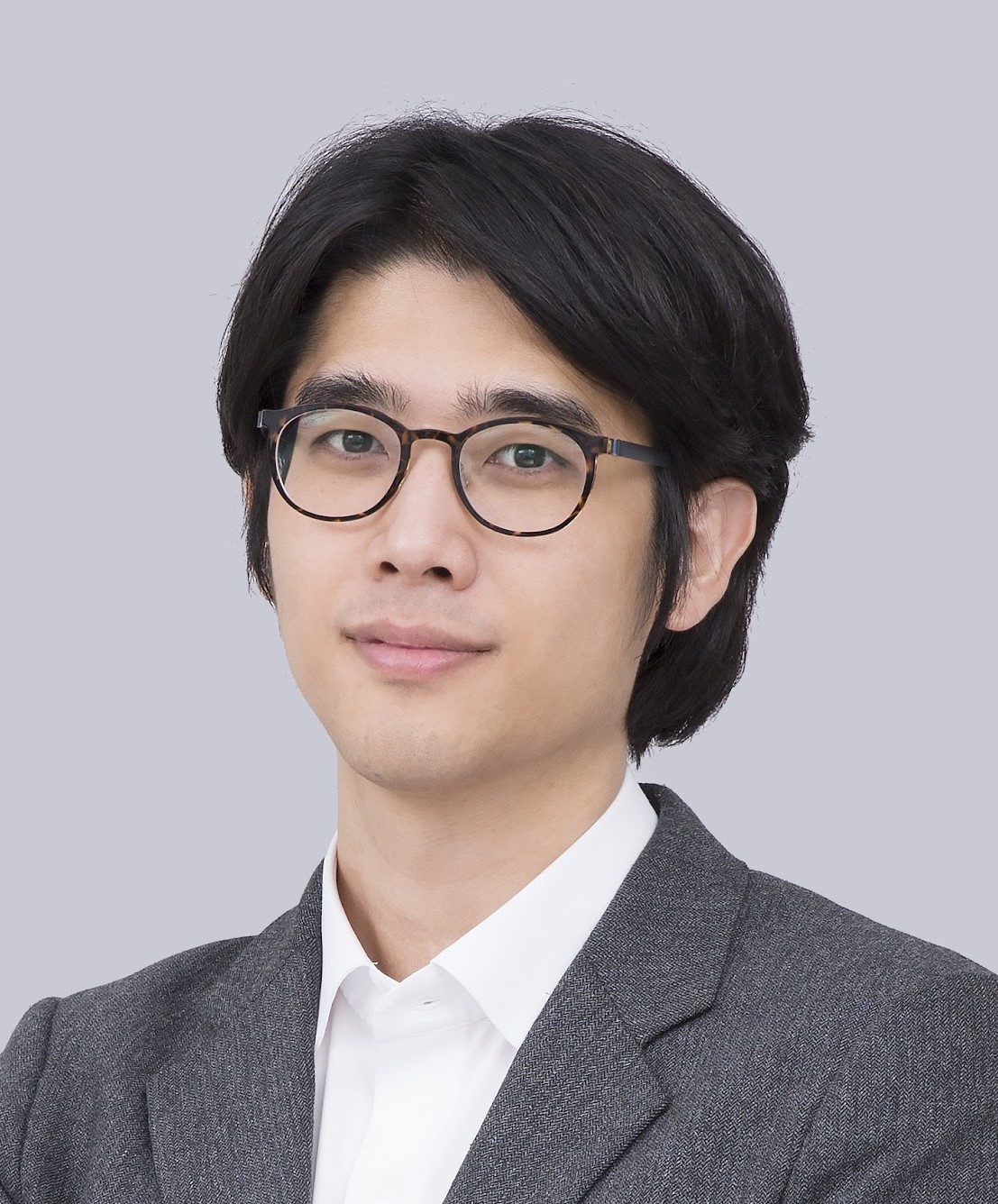}}]{Jihong Park} 
(Senior Member, IEEE) is a Lecturer with the School of Information Technology, Deakin University, Australia. He received his B.S. and Ph.D. degrees from Yonsei University, Seoul, South Korea, in 2009 and 2016, respectively. He was a Postdoctoral Researcher at Aalborg University, Denmark, from 2016 to 2017, and the University of Oulu, Finland, from 2018 to 2019. His current research interests include AI-native and semantic communications, as well as distributed and quantum machine learning. He is a Member of ACM and AAAI. He has served as a Program Committee Member for several conferences and workshops, including IEEE GLOBECOM, ICC, and INFOCOM, as well as NeurIPS, ICML, and IJCAI. He has organized workshops at IEEE GLOBECOM, WCNC, VTC, and SECON. He has received 2023 IEEE Communication Society Heinrich Hertz Award, 2022 FL-IJCAI Best Student Paper Award, 2014 IEEE GLOBECOM Student Travel Grant, 2014 IEEE Seoul Section Student Paper Prize, and 2014 IDIS-ETNEWS Paper Award. Currently, he is the co-chair for 2023 IEEE GLOBECOM Symposium on Machine Learning for Communications, and an Associate Editor of Data Science for Communications (Frontiers) and Signal Processing for Communications (Frontiers).
\end{IEEEbiography}

\begin{IEEEbiography}[{\includegraphics[width=1.1in,height=1.3in, clip,keepaspectratio]{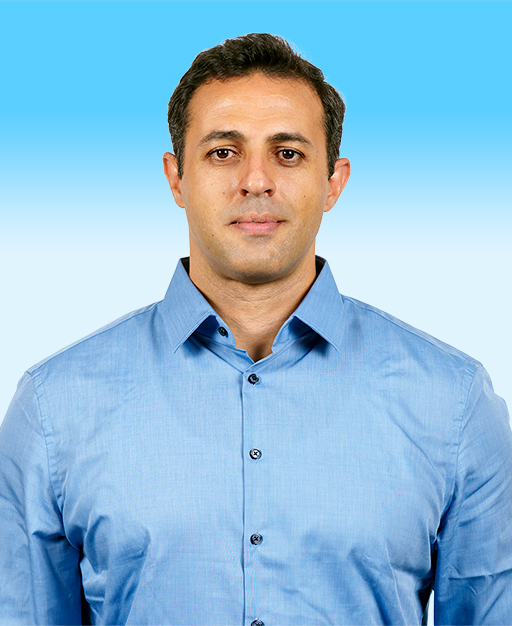}}]{Mohamed Wahib} is a team leader of the “High-Performance Artificial Intelligence Systems Research Team” at RIKEN Center for Computational Science (R-CCS), Kobe, Japan. Prior to that he worked as is a senior scientist at AIST/TokyoTech Open Innovation Laboratory, Tokyo, Japan. He received his Ph.D. in Computer Science in 2012 from Hokkaido University, Japan. His research interests revolve around the central topic of high-performance programming systems, in the context of HPC and AI. He is actively working on several projects including high-level frameworks for programming traditional scientific applications, as well as high-performance AI.
\end{IEEEbiography}

\end{document}